\documentclass[aps,onecolumn,prb,tightenlines,floatfix,showpacs]{revtex4}
\usepackage{graphicx}
\usepackage[english]{babel}
\usepackage{amsmath}
\usepackage{amssymb}
\usepackage{times}
\usepackage{verbatim}

\newcommand{\vect}[1] {\mathbf{#1}}

\newcommand{\uk}{u_{\mathbf{k}}}
\newcommand{\vk}{v_{\mathbf{k}}}

\newcommand{\up} {\uparrow}
\newcommand{\down} {\downarrow}

\def\beq{\begin{eqnarray}} \def\eeq{\end{eqnarray}}
 \def\Re {\mbox{Re}} \def\Im {\mbox{Im}}
  
\def\up{\uparrow} \def\down{\downarrow} %\def\G{\mathcal{G}}
 \def\c{\hspace{2pt}}

\begin{document}
\title{Perfect Fluids and Bad Metals: Transport Analogies Between
Ultracold Fermi Gases and High $T_c$ Superconductors}

\author{Hao Guo$^{1}$, Dan Wulin$^{2}$,
Chih-Chun Chien$^{3}$ and K. Levin$^{2}$}
\affiliation{$^1$Department of Physics, University of Hong Kong, Hong Kong, China}
\affiliation{$^2$James Franck Institute and Department of Physics,
University of Chicago, Chicago, Illinois 60637, USA}

\affiliation{$^3$Theoretical Division, Los Alamos National Laboratory, MS B213, Los Alamos, NM 87545, USA}

\pacs{03.75.Ss,67.10.Jn, 67.85.De}

\date{\today}
\begin{abstract}
In this paper, we examine in a unified fashion dissipative
transport in strongly correlated systems. We thereby demonstrate
the connection between ``bad metals"
(such as the high temperature superconductors) and ``perfect fluids" 
(such as the ultracold Fermi gases, near unitarity).
One motivation of this work is to communicate to the
high energy physics community some of the central unsolved
problems in high $T_c$ superconductors. Because of interest
in the nearly perfect fluidity of the cold gases and
because of
new tools such
as the AdS/CFT correspondence, this better communication may
lead to important progress in a variety of different fields.
A second motivation is to draw attention to the great power
of transport measurements which more directly reflect 
the excitation spectrum than, say, thermodynamics
and thus strongly
constrain microscopic theories of correlated fermionic superfluids.
Our calculations show that bad metal and
perfect fluid behavior is associated with the presence of a normal
state excitation gap which suppresses the effective number
of carriers leading to anomalously low conductivity and viscosity
above the transition temperature $T_c$. Below $T_c$ we demonstrate that
the condensate collective modes (``phonons")
do not couple to transverse probes such as the shear
viscosity. As a result, our calculated shear viscosity at low $T$ becomes
arbitrarily small as observed in experiments.  In both homogeneous and
trap
calculations
we do not find
the upturn in $\eta$ or $\eta/s$ (where $s$ is the
entropy density) found in
most theories. 
In the process of these studies we demonstrate compatibility with
the transverse sum rule and find reasonable
agreement with both viscosity and cuprate conductivity
experiments.
\end{abstract}

\maketitle

\section{Introduction}
\label{sec:1}

There is a widespread interest in studying ultracold Fermi
gases to learn about some of the most strongly interacting
systems in nature.
It has been argued \cite{Physicstoday}
that the atomic Fermi gas near unitarity provides
a prototype for the physics of quark-gluon plasmas which are associated
with the early stages of the Big Bang.
Most remarkable about these two nominally different physical systems is
the fact that they correspond to nearly perfect fluids, exhibiting
even in their non-superfluid phases extremely small values of
the shear viscosity $\eta$. In this paper we call attention to
a third class of strongly interacting systems which is the metallic
counterpart of the perfect (neutral) fluid, namely,
the so-called ``bad metal'' \cite{EmeryKivelsonPRL74}.
We show how low conductivity and low viscosity
are
analogous.
Bad metals are known to exist in nature in a most fascinating
class of materials: the high $T_c$ superconductors. We argue here that
to learn more about the physics of nearly perfect fluids
it is particularly useful to study the behavior of the
conductivity in the cuprate superconductors and vice versa.
We note that it may seem at first sight paradoxical that 
bad metals and
perfect fluids have anything in common since the condensate
contribution to the conductivity
is infinite in the metallic system whereas
the viscosity counterpart in the neutral superfluid is zero.
We stress throughout this paper that the dc transport 
(associated with $\omega \rightarrow 0$) in both cases reflects the
normal fluid or \textit{excitations} of the condensate.

Because the ultracold Fermi gases near the unitary limit are also thought to
be related to quark-gluon plasmas \cite{Physicstoday}, 
much attention
has focused on particle-physics-based calculations of the anomalously
low shear viscosity, $\eta$ \cite{Son1,Rupakvisc}. However, considerable
insight into the thermodynamics \cite{ThermoScience} and
various spectroscopic studies \cite{Ourreview,RFReview} has
also been obtained via a condensed matter perspective.
This paper belongs to the second school in which
BCS theory is extended to accommodate arbitrarily strong interactions.
We apply
BCS to Bose Einstein condensation (BEC) crossover
theory \cite{Ourreview} to compute
$\eta$ in the neutral Fermi gases and
the conductivity
$\sigma$
(evaluated at $\omega \rightarrow 0$)
for the charged counterparts, such as the high $T_c$ superconductors. Our
results, which are reasonably consistent with experiments, apply both
above and below $T_c$. 
Essential to this work is
demonstrable consistency with central
sum rules and conservation laws \cite{KadanoffMartin}.

Experimentally this crossover
can be straightforwardly observed in the ultracold Fermi gases
by exploiting Feshbach resonances.
As has been argued \cite{LeggettNature}, this crossover also appears
relevant to the cuprates (i) because
of their
anomalously short
coherence length.  Added support comes from
(ii) their anomalously high transition temperatures and
(iii) the widespread effects of a normal state gap
\cite{Ourreview,Jin6}.
%DAN CHANGE
(iv) The smooth evolution from over to underdoped behavior of
the optical conductivity, has similarly been used to argue \cite{Deutscher2}
that the BCS-BEC crossover scenario might be appropriate to the cuprates.
Although the cold gases and high $T_c$ cuprates
have $s$ and $d$-wave order parameters respectively,
this difference does not significantly modify our treatment of transport.
%For simplicity in our theoretical discussions we present our
%few equations for
%the $s$-wave case.

The shear viscosity, like the dc conductivity,
is a powerful probe for testing microscopic
theories because it reflects the
normal fluid component and is extremely sensitive
to the nature of the excitation spectrum.
The low $T$ normal Fermi liquid phase with scattering
lifetime $\gamma^{-1}$ and effective mass $m^{\ast}$
helps to set up the general
nomenclature. Here $\eta=\frac{1}{5}nv^2_F\gamma^{-1} m^{\ast}$.
Similarly $\sigma = \frac{n e^2}{m^{\ast}} \gamma^{-1}$.
More generally, one can think of $\eta$ and $\sigma$
as characterized
by the effective number of
normal excitations ($n \rightarrow n_{eff}(T)$)
as well as their lifetime which we emphasize here is
a many body effect.
Crucial is an understanding of
how $n_{eff}$ depends on $T$.

Our central conclusion is that both the effects of a
fermionic gap (with onset temperature $T^{\ast} > T_c$) and 
non-condensed pairs
act in concert to reduce $n_{eff}$ and thus
\textit{lower} the shear viscosity
and dc conductivity at all $T < T^{\ast}$.
These non-condensed pairs are associated with the stronger than
BCS attraction and are present for $0<T<T^{\ast}$.
When quantitatively compared with
very recent shear viscosity experiments \cite{ThomasViscosityScience_online}
(we independently infer \cite{OurComparison}
an estimated lifetime from
radio frequency data) the agreement is reasonable.
%With a lifetime
%so constrained,
%there are no free parameters.
%Using previous thermodynamical
%calculations \cite{ThermoScience}
%of the trap energy $E$ and entropy density
%$s$, a plot of the trap-integrated $\eta$ decreases
%with decreasing $E$, while $\eta/s$ appears to be roughly constant
%at low $E$, but above the
%universal quantum limit \cite{Son1}.
We similarly compare the ($\omega \rightarrow 0$) dc conductivity (
as well as its normal state
$\omega \equiv 0$ inverse, $\sigma_{dc}^{-1}(T) = \rho(T)$, the resistivity) with the
counterparts in the high $T_c$ superconductors and find reasonable agreement 
with trends in doping and temperature.
While condensed matter ``simulations" based on 
these atomic Fermi gases 
are now of wide interest, our approach here is somewhat
different because we 
focus on condensed matter phenomena which
currently exist in nature, for example the pseudogap
in the cuprate superconductors.
We have emphasized the commonality for these
two systems in past reviews \cite{Ourreview}
and more recently \cite{RFReview} in the context of analogous tools such as
radio frequency spectroscopy, which is closely related to photoemission \cite{Jin6}. 
It seems natural then to investigate here the analogies in transport.

Our work is organized around a second important
premise: any theory of transport
or scattering in the superfluid Fermi gases (both above and below $T_c$)
must be formulated in a fully gauge invariant and sum-rule consistent fashion. One could imagine doing
more and more sophisticated many body theories than standard BCS-related
approaches (i.e. summing more and more diagrams) but, if sum rules and conservation laws cannot be
demonstrated to hold, we would
argue the significance of these calculations is problematic.
Indeed,
a major theme of our recent \cite{OurBraggPRL,Ourviscosity}
work which is summarized in this
overview is to arrive at a demonstrably consistent theory
of this exotic superfluidity by extending the important
contributions of Nambu \cite{Nambu60}
to the case of BCS-BEC crossover.
This will serve as a basis for
computing gauge invariant and properly conserving response functions which are measured
in transport and scattering experiments.
It is not as straightforward to demonstrate this consistency in
Monte Carlo or other numerically-based schemes. Moreover, as theories
of BCS-BEC crossover become more and more diagrammatically complicated
it is not straightforward to determine that they are
compatible with gauge invariance and conservation laws.
Once this compatibility is established the next step is to
assess 
the physical implications of this consistency in experimental probes.
We stress here that our approach to BCS-BEC crossover is semi-analytical
and therefore reasonably transparent.  Moreover, unlike alternative (more numerical
diagrammatic schemes \cite{Ourcompare}) the normal to superfluid transition
is properly smooth and second order.

A systematic treatment of transport in the
Fermi superfluids requires the introduction \cite{Regge77}
of inter-dependent fermionic as well as bosonic (or pair) excitations.
In past transport literature there has been a focus on either one
\cite{Bruunviscosity} or the other \cite{Rupakvisc},
but not both.
Here we use a Kubo-based formalism which readily accommodates the
simultaneous bosonic and fermionic excitations of the
normal state and condensate and thereby
addresses $n_{eff}$
quite accurately
while the
alternative Boltzmann or kinetic theory-based approaches
do not naturally incorporate these multiple
statistical effects.
The Kubo approach
includes scattering processes via
the lifetimes \cite{KadanoffMartin2} which appear in
the various Green's functions, while Boltzmann schemes treat
lifetimes via collision integrals.  However, because the physics of this
dissipation is principally associated with the many body
processes of boson-fermion inter-conversion (via a parameter
$\gamma$ which appears throughout this paper),
it should be satisfactorily addressed
only in theories which treat the mixed statistics.

We stress in this paper that dissipative transport in
the normal phase must also be addressed if one is to fully understand
the counterpart below $T_c$ case.
Except in strict BCS theory, this normal phase is affected by the below $T_c$
pairing correlations. As a result, consistency checks need to be applied
above $T_c$ and the behavior must necessarily reflect that the transition is second order.
Our central theme is
based on the fact that the fermionic excitation gap or pairing gap persists above
$T_c$ as a ``pseudogap" (except in strict BCS theory); this leads to important
transport implications such as ``nearly perfect fluidity" (reflecting
anomalously small viscosity \cite{Physicstoday}) or in
the analogous charged system ``bad metal" \cite{EmeryKivelsonPRL74}
behavior (reflecting anomalously
small conductivity).

To add to the case for simultaneously studying both the viscosity and dc conductivity
we note that the
wealth of data available \cite{TimuskRMP} and the relative ease of
measurement (compared to those in atomic or
RHIC experiments) make the cuprates a particularly useful analogue
for the nearly perfect fluids. In addition to dc measurements, one
can probe \cite{TimuskRMP} the conductivity as a function of frequency
$\omega$, $\sigma_{ac}(\omega)$, over a wide range of $\omega$. 
Because of the existence of a frequency sum rule,
theories of the (dc) conductivity are thereby highly
constrained.
Importantly, this sum rule
and an in depth understanding of the conductivity serve to constrain
analogous microscopic theories of viscosity.
Conventionally, 
``bad'' metals are systems
in which the estimated mean free path 
$l$ is
shorter than all length scales; along with anomalously low
conductivity this leads to the absence of resistivity
saturation. The descriptive ``perfect'' is also associated with a situation in
which $l$ is small compared to physical length scales
\cite{Weinberg}.
In strongly correlated superfluids, we re-iterate here that 
small $l$ does not solely reflect short transport lifetimes $\tau$ but
rather a notable suppression in the effective ``carrier number''. 
The
influence of bad metallicity
on superconductivity was studied in Ref.~\onlinecite{EmeryKivelsonPRL74}. Here we emphasize
the converse: the influence of superconductivity on transport \cite{AndoReview}. Moreover,
as will be discussed in
more detail below, pairing fluctuations and the phase fluctuations invoked 
earlier \cite{EmeryKivelsonPRL74} are clearly
distinct.

We have included a summary of the main results of this paper in Section
\ref{sec:1C}. 
For ease in reading this paper, the reader who is not interested in
the technical details can skip Sections \ref{sec:2} and \ref{sec:3}
and go directly to Section \ref{sec:4}.

\subsection{Experimental Overview}
\label{sec:1A}

\begin{figure*}
\includegraphics[width=1.3in,clip]
{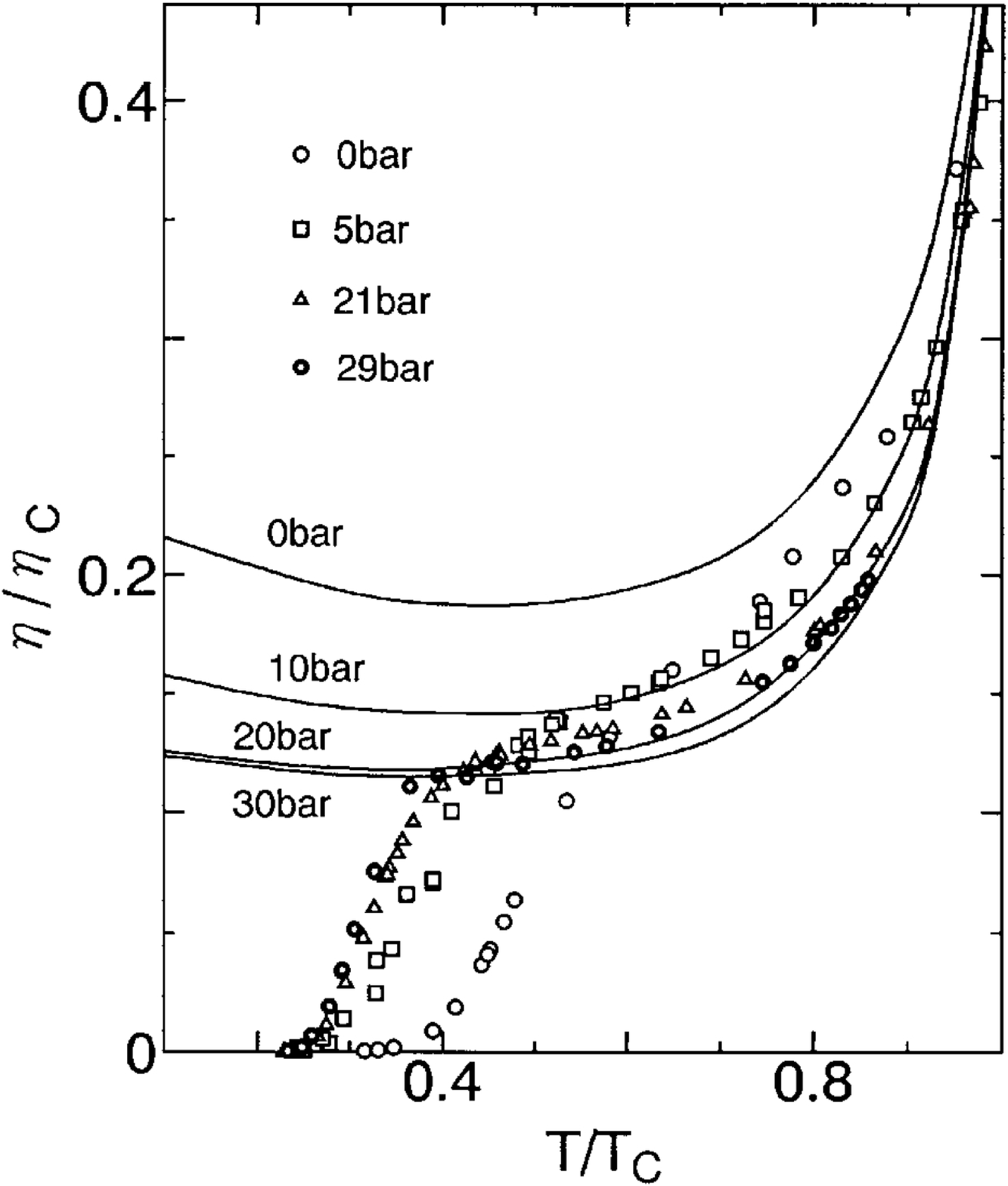}
\includegraphics[width=1.6in,clip]
{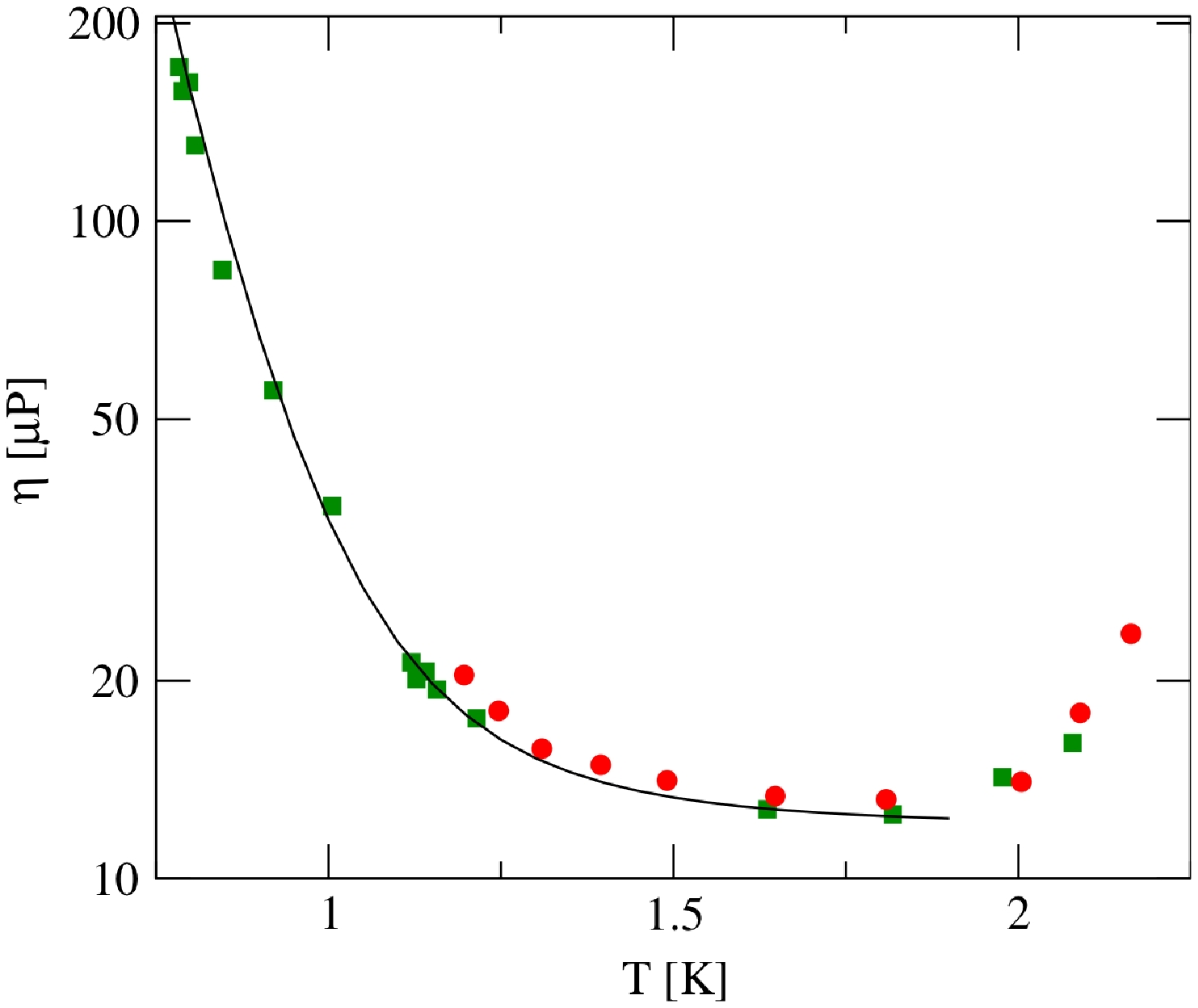}
\caption{Viscosity in helium-3 (left) from Ref.\onlinecite{Roobol}
and helium-4 from Ref.\onlinecite{Woods} (right).
The solid curves on the left represent theoretical predictions
which were not verified. The viscosity vanishes for low temperature in the case of helium-3 because of a suppression of fermionic excitations. The
behavior of $\eta$ in helium-4 is governed by the rotons and
phonons.}
\label{fig:1}
\end{figure*}

To understand viscosity in fermionic superfluids, it is useful
to begin with helium-3
which has been successfully described
using BCS-based approaches, albeit for the $p$-wave case 
\cite{Viscositypapers,Dorfle80}.
Here experiments \cite{Helium3} indicate that $\eta$
drops off rapidly to zero in the superfluid phase.
This is shown in Figure \ref{fig:1} to the left. Interestingly, there is a
minimum in $\eta$ \textit{above} $T_c$ which is associated with strict
Fermi liquid behavior \cite{VollhardtRMP}. In a Fermi liquid the
number of carriers and mass are both $T$ independent,
while the inter-fermion scattering lifetime
varies as $T^{-2}$.
The standard interpretation of the data below $T_c$ is that $\eta$ decreases with
decreasing $T$ as a result of
the suppression of fermionic excitations at low $T$. 
In a strong magnetic field one spin 
component of the triplet is driven normal and this leads to a
very different behavior for the shear viscosity \cite{Roobol},
in which (even below $T_c$) it reflects the normal Fermi liquid behavior
above $T_c$. In this $A_1$ phase, the low temperature behavior
exhibits an upturn at low $T$; this
is not to be associated with coupling to collective modes
or phonons, but rather reflects a residual normal component. 
In BCS-based superfluids, we stress \cite{Nambu60} that
Nambu-Goldstone
boson effects
do not naturally enter into the
transverse transport properties such as $\eta$.
By contrast, in
the helium-4 counterpart shown on the right in Fig. \ref{fig:1}, 
the single particle bosonic
excitations couple to the collective (Nambu-Goldstone) modes,
leading to an upturn \cite{Woods} in $\eta$ at low $T$, which has
also been predicted (but not seen)
for the atomic Fermi superfluids \cite{Rupakvisc}.

\begin{figure}
\includegraphics[width=1.8in,clip]
{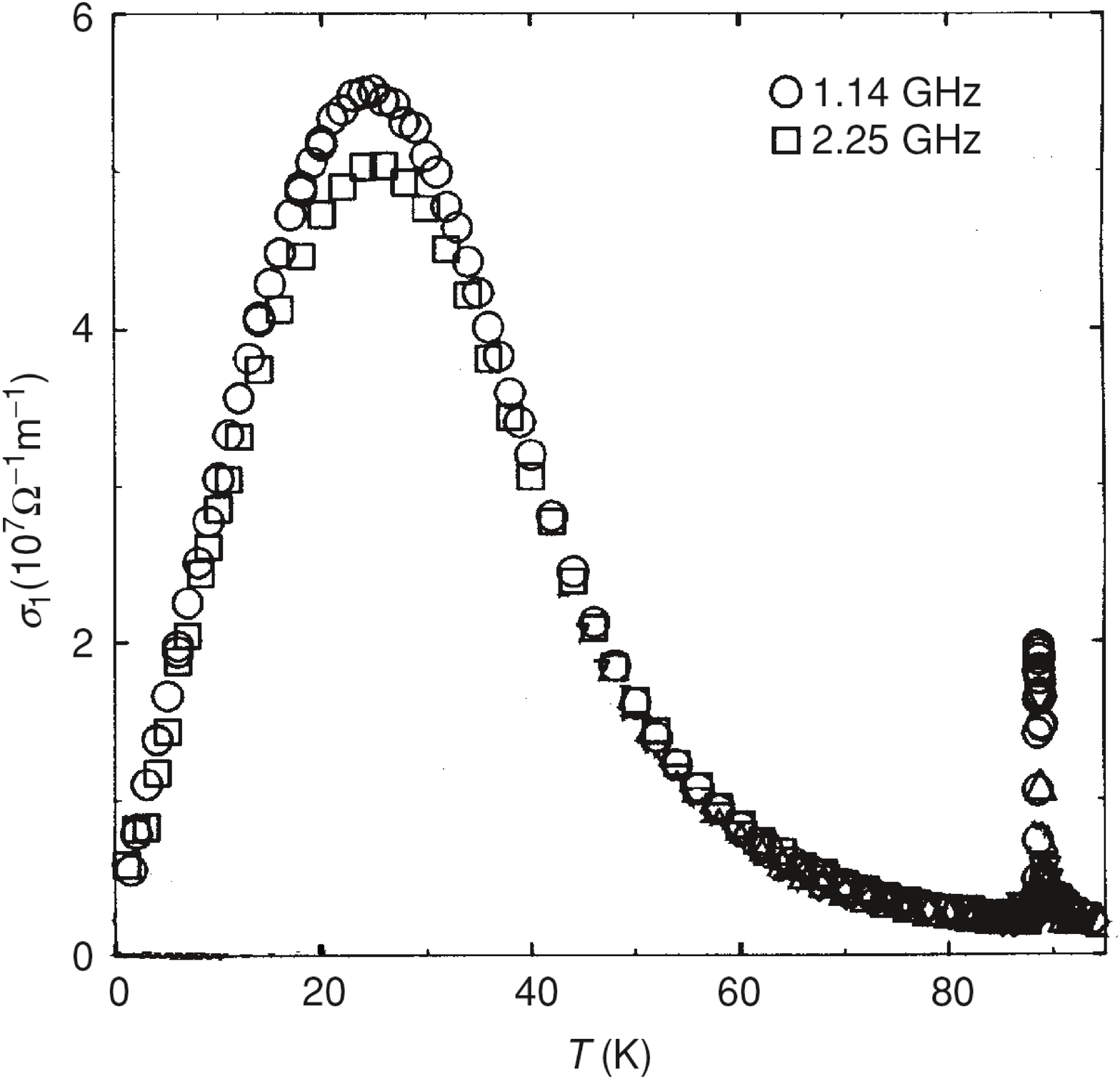}
\includegraphics[width=1.3in,clip]
{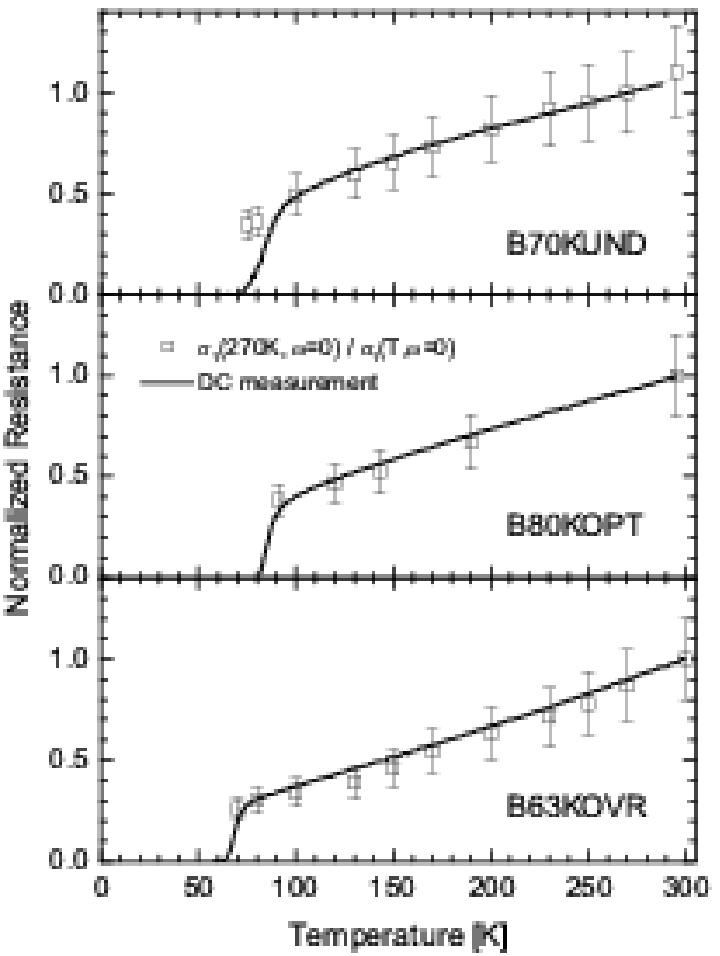}
\caption{Temperature dependence of  
$\omega \rightarrow 0$ conductivity \cite{Orenstein} (left)
and resistivity \cite{Bontemps} (right)
in the cuprates. The conductivity exhibits a maximum as a function of temperature because of competition between the effective carrier number and quasiparticle lifetime.  Increasing the doping tends to lower the resistivity.}
\label{fig:3}
\end{figure}

For strongly correlated charged systems, the counterpart experiments are summarized
in Figure \ref{fig:3}, here for the high $T_c$ cuprates, which have
$d$-wave pairing. On the left is the low $T$, $\omega \rightarrow 0$
conductivity for a $T_c \approx 90K$ sample. This shows the
fairly generic maximum below $T_c$. 
The figure on the right is the resistivity or inverse conductivity for
different stoichiometries as a function of
temperature. Above $T_c$, two crucial points are that
$\rho = \sigma^{-1}$ is nearly linear with $T$ and its magnitude
seems to decrease from the UND (underdoped) to the OVR (overdoped) samples.
It is inferred \cite{Timusk}
from similar systematic studies that the
effective number of carriers at fixed $T$ is
substantially depressed, varying as the doping $x$ rather
than the expected $1+x$.
Importantly, for the present purposes
$\sigma(T)$ seems to approach zero at the lowest temperatures.  The latter point is
consistent with the vanishing $\eta$ shown in the previous
figure for the fermionic superfluid. The pronounced maximum is
thought to arise from the competition between the decrease in
$n_{\rm{eff}}$ and the increase in the fermionic lifetime
at low $T$. Such a competition is not nearly as apparent in an
$s$-wave system where $n_{\rm{eff}}$ is exponentially suppressed.

\begin{figure*}
\includegraphics[width=1.8in,clip]
{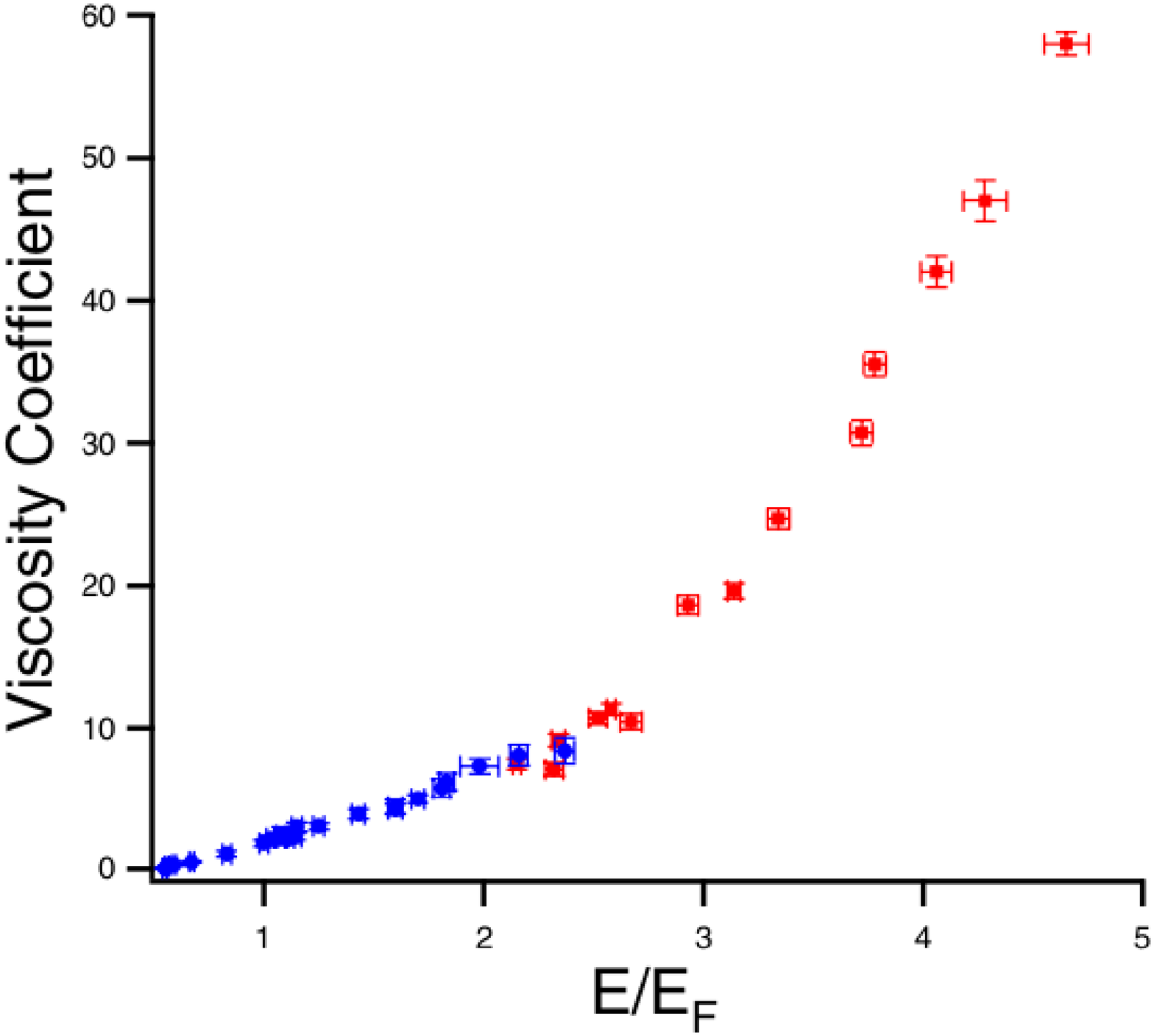}
\includegraphics[width=1.8in,clip]
{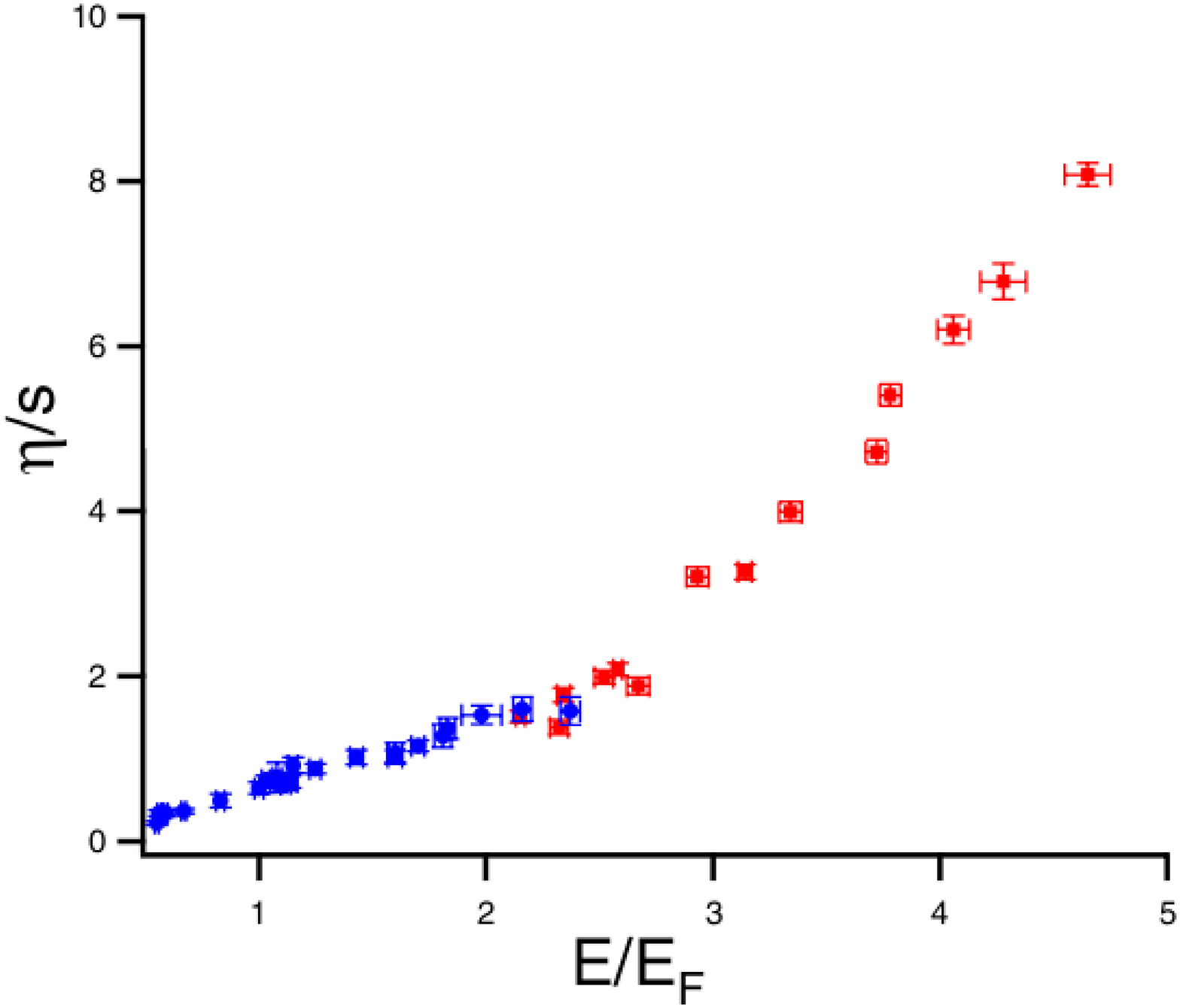}
\caption{Viscosity experiments on unitary Fermi gases from recent experimental results showing the shear viscosity $\eta$ and the ratio of viscosity and entropy density $\eta/s$, 
Ref.\onlinecite{ThomasViscosityScience_online}. Importantly, the viscosity vanishes are low temperature and this puts a strong constraint on the theories of the unitary Fermi gas.}
\label{fig:2}
\end{figure*}

In Fig. \ref{fig:2} we summarize recent viscosity experiments from the Duke
group
\cite{ThomasViscosityScience_online} on a unitary trapped
Fermi gas. One sees here that despite previous predictions,
the viscosity and its ratio to entropy density are both
strongly suppressed at low $T$.  
We will return to all of these figures throughout the paper.

\subsection{Theoretical Overview}
\label{sec:1B}

The BCS-BEC crossover theory that is adopted here is based on a natural
extension of the BCS ground state,
\begin{equation}
\Psi_0=\Pi_{\bf k}(\uk+\vk
%c_k^{\dagger} c_{-k}^{\dagger})
c_{\vect k\up}^\dagger c_{-\vect k\down} ^\dagger) |0\rangle .
\label{eq:GS}
\end{equation}
This ground state is often called the ``BCS-Leggett''
state.
The observation that this state is much more general than originally
presumed forms the basis for BCS-BEC crossover theory
\cite{Leggett,Eagles,NSR}. To implement this generalization of the
BCS ground state, all that is required is that one solve for the two
variational parameters $\uk$ and $\vk$ in concert with a self
consistent condition on the fermionic chemical potential $\mu$. As
the attraction is increased, $\mu$ becomes different from the Fermi
energy $E_F$. The
two variational parameters $\uk$ and $\vk$ can be
converted to two more physically accessible parameters associated
with the zero temperature gap (or equivalently order parameter) and
$\mu$. We stress that while in the cuprates the system is far from
the BEC limit (in large part because of $d$-wave lattice effects
\cite{Chienlattice2}), it is nevertheless quite distinct from the
weak-coupling BCS limit due to the anomalously short coherence
length \cite{LeggettNature}.

Since this wave function is so closely related to the BCS state, it
is natural to ask whether its behavior away from $T=0$ can be
consolidated into as simple a formalism and physical picture as
there is for the ground state. In Section \ref{sec:2}, we answer this
question in the affirmative by recasting the equations of
conventional BCS theory using a formalism that can then be readily
generalized to include BCS-BEC crossover.

\subsubsection{Simple Physical Picture of BCS-BEC Crossover Scenario}
\label{sec:1B1}

\begin{figure*}[t]
\includegraphics[width=5.8in,clip]
{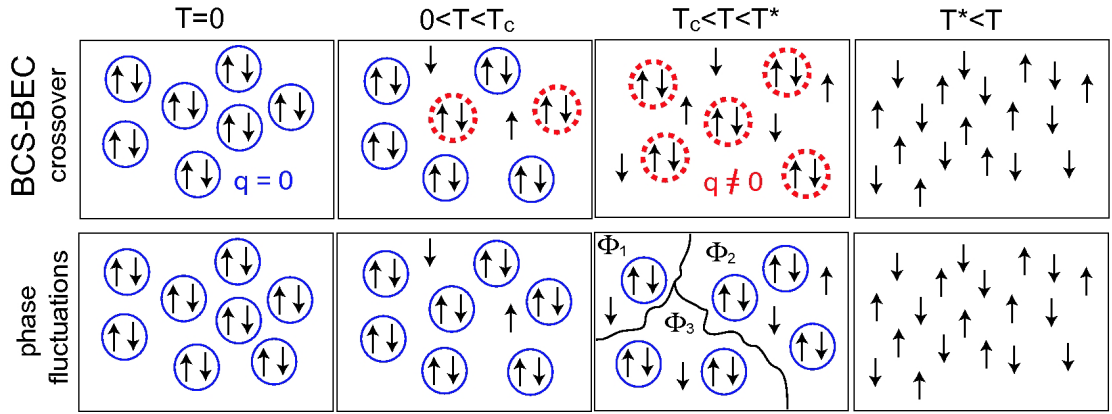}
%{theorycomp.eps}
%{Oldcartoon.eps}
\caption{\label{fig:4} (Color online) Cartoon comparing the
BCS-BEC crossover and phase fluctuation scenarios. Throughout, blue
closed circles, lone arrows, and dashed red circles represent
condensed fermion pairs, unpaired fermions, and finite lifetime
pairs, respectively. The crossover theory is distinguished by the
presence of noncondensed $\textbf{q}\neq 0$ pairs for nonzero
temperatures less than $T^*$. The defining feature of the phase
fluctuation picture is the presence of different phase domains above
$T_c$, indicated by the regions labelled with different $\Phi$'s.}
\end{figure*}

Before introducing the $T \neq 0$ formalism,
we present a simple picture of the excitation spectra
which ultimately enter into transport.
The top row in Fig. \ref{fig:4} (from left to right) shows the
schematic behavior as one passes from the $T=0$ BCS-Leggett
ground state to the above $T^*$ Fermi liquid. The red (dotted
circles) pairs are associated with net finite momentum, while the
blue (solid circles) pairs correspond to the phase coherent
condensate with zero center of mass momentum and the lone arrows
represent fermionic excitations. The first panel shows that the
ground state consists of fully condensed pairs [as in
Eq.~(\ref{eq:GS})], while the second panel shows that below
$T_c$ but above $T = 0$ there are both condensed and non-condensed
pairs along with fermionic excitations. These non-condensed pairs
persist above $T_c$ (third panel) in the form of ``preformed" pairs,
while the condensed pairs are no longer present. Finally at
temperatures above $T^*$ all bosonic-like excitations are absent;
the only excitations are fermionic. The second panel with $0<T<T_c$
is the most interesting from the perspective of the present paper.
In the cuprates, this is the regime in which the widely discussed \cite{ShenNature}
''two-gap`` physics appears. Here the
coexistence of the condensate and of non-condensed pairs, leads to
two gap contributions \cite{Kosztin1}, one associated with the
pseudogap (called $\Delta_{pg}$) and another associated with the
condensate (called $\Delta_{sc}$).

In contrast to this pairing
fluctuation picture of the BCS-BEC crossover theory stands the phase
fluctuation picture shown in the bottom row of Fig. \ref{fig:4}.
In this scenario of the pseudogap \cite{Emery} there exist finite
size regions of superconducting order in the normal state. While the
amplitude of the pairing gap $\Delta_{\bf k}$ is fixed in these
regions, the superconducting phase, labeled by $\Phi$, fluctuates
spatially between them. The degree of phase fluctuation increases
with temperature, and the Fermi liquid state is reached at $T>T^*$.
In contrast, the phase is long-range ordered below $T_c$ and the
superconducting phase is described by the BCS state consisting 
only of
condensed pairs and unpaired fermions. In this picture there appears
as yet to be no counterpart to the ``two gap physics" below $T_c$ seen
in the crossover scenario.
In a simplified fashion, it
could be said that the preformed pair theory is concerned with
fluctuations in momentum space (in terms of $\textbf{q}\neq0$
pairing) while the phase fluctuation picture focuses on fluctuations
of the phase in real space.

In the BCS-BEC crossover scenario, the effects of condensed [i.e.,
superconducting (sc)] and non-condensed [i.e., pseudo-gap (pg)]
pairs are described by two distinct contributions to the fermionic
self energy $\Sigma({\bf k}, \omega)$ 
\begin{eqnarray}\label{eq:selfenergy}
\Sigma({\bf k},\omega) = \Sigma_{pg,{\bf k}} + \Sigma_{sc,{\bf k}}  
= \frac{\Delta_{pg,\bf k}^2}{\omega+\xi_{\textbf{k}}+i\gamma} +
\frac{\Delta_{sc,\bf k}^2 }{\omega+\xi_{\textbf{k}}} \ .
\end{eqnarray}
The gap functions $\Delta_{pg,{\bf k}}$ and $\Delta_{sc,{\bf k}}$
are assumed to follow either a simple $s$ or $d$-wave form.
The condensed
pairs have the usual BCS self energy contribution, $\Sigma_{sc}$,
while the self energy of the non-condensed pairs $\Sigma_{pg}$
possesses an additional term, $\gamma$, with $\gamma^{-1}$
reflecting the finite lifetime of the non-condensed pairs. This
form of $\Sigma_{pg}$ was derived microscopically in
Ref.~\onlinecite{MalyThesis} using a $T$-matrix approach (see below).
It plays a central role in transport, largely through Ward
identities which relate the self energy to transport
properties.
At the microscopic level, it is important to stress
that the above expression for $\Sigma_{pg}$ is not generic to all $T$-matrix
theories, but strongly depends on an underlying BCS-like structure
of the ground state associated with the present approach.

\subsubsection{Lifetime Effects}
\label{sec:1B2}

In our discussion of dissipative transport, a crucial point is to address
the origin of finite lifetime effects.
Throughout this paper we will argue that \textit{the central dissipation
process is associated with the inter-conversion from fermions to pairs}.
We stress that this is a many body effect 
and should not be associated with the two body scattering length.
Nevertheless,
at unitarity, these inter-conversion processes are likely
to lead to the shortest lifetimes simply because the 
number of fermions and bosons is roughly equal there. 
This is in contrast to the BCS (BEC) regime in
which there are virtually no bosons (fermions).
This physical picture is consistent with
Eq.~(\ref{eq:selfenergy})
which above $T_c$ has been rather widely adopted by 
the high $T_c$ community \cite{Normanphenom,Normanarcs,FermiArcs} and the cold
Fermi gas community \cite{Jin6}. In this second context, it is
this form of the fermionic self energy (or equivalently spectral
function) which is to be associated with the downward dispersing
quasi-particles revealed in momentum resolved radio frequency
experiments \cite{momentumRF,RFReview}.
In the cuprates,
above $T_c$, it is
the 
finite lifetime of the non-condensed pairs which leads to the
interesting physics
associated with the ``Fermi arcs". These have been
interpreted as a blurring of
the $d$-wave nodes. Importantly, below $T_c$ one sees their sudden collapse to
conventional point nodes \cite{FermiArcs,Kanigelarcs} as a result
of the onset of the order parameter $\Delta_{sc}$.

\subsection{Central Results of this Paper}
\label{sec:1C}
\begin{figure*}
\includegraphics[width=4.5in,clip]
{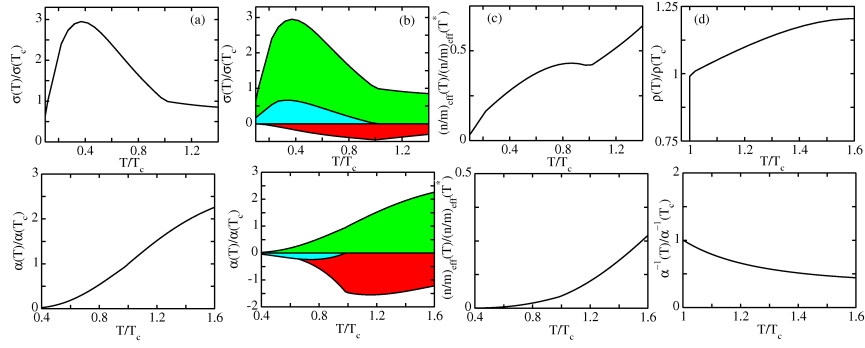}
%{vis_weak_displayfinal4.eps}
%{Haodelta_display.eps}
%\includegraphics[width=4.5in,clip]
%{con_weak_displayfinal2.eps}
\caption{ Summary figure of $d$-wave conductivity (top 4 panels)
and $s$-wave viscosity (bottom 4 panels) as a function of temperature.
Both panels (a)
and the resistivity at the top in panel (d) (or inverse conductivity)
can be compared with experiments in Figures \ref{fig:2}
and \ref{fig:3}.
Panels (b) are contributions to
transport from 3 components. Here
red = pg, blue = sc, while
green reflects the
difference to make up the total. Panels (c) plot the
effective carrier number and (d) the inverse transport
coefficients.
}
\label{fig:13}
\end{figure*}

\begin{figure}
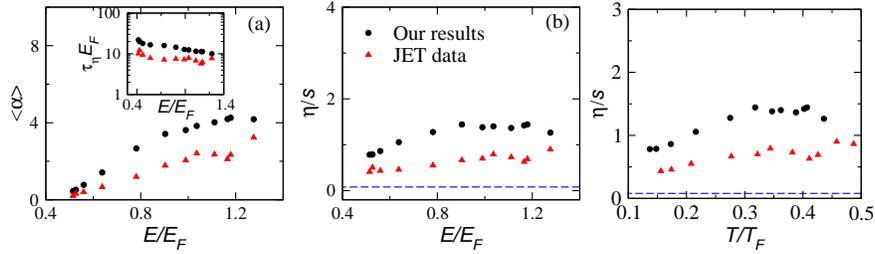

\begin{center}
\includegraphics[width=3.in,clip]
{JET2_our_trap.eps}
\includegraphics[width=1.5in,clip]
{HaoCompare15.eps}
\caption{ (a) Comparison of shear viscosity
$\eta \equiv \alpha
n\hbar$
and experiments \cite{ThomasJLTP08} (red triangles) at unitarity
for a trapped gas. In theory
plots (black dots) we use the
calculated thermodynamics for the trap energy $E$ and entropy density
$s$.
The inset in (a) plots the estimated transport lifetime from radio frequency
experiments \cite{OurComparison},
as compared with the experimentally deduced lifetime needed for
an exact fit to the theory. (b) Comparison of $\eta/s$. The blue dashed line labels
the quantum lower
limit of $\eta/s$ given by string theory\cite{Son1}.
The last panel on the right converts the horizontal axis in (b) to
temperature.
}
\label{fig:12}
\end{center}
\end{figure}

\subsubsection{Summary of Central Formulae}
\label{sec:1C1}

In the weak dissipation
limit (where $\gamma$ is small), we may write down a combined expression for the shear viscosity
and dc conductivity.
We have 
\begin{widetext}
\begin{eqnarray}
     \begin{array}{l}
       \eta\\
       \sigma
     \end{array}
   \Bigg\}
=c_{\eta,\sigma}\displaystyle{\int_0^{\infty}}dp\frac{p^{6,4}}{m^2}
\frac{E^2_{\bf p}-\Delta_{\textrm{pg}}^2}{E^2_{\bf p}}
(u^2_{\bf p}\mp v^2_{\bf p})^2(-\frac{\partial f(E_{\bf p})}{\partial E_{\bf p}})
%\displaystyle{\lim_{\gamma_{\eta,\sigma}\rightarrow0}}
\tau_{\eta,\sigma}
\label{eq:19}
\end{eqnarray}
\end{widetext}
Here $E_{\bf p}$ is the usual BCS dispersion,
$E_{\bf p} \equiv \sqrt{\xi_{\text{p}} + \Delta^2}$
where $\Delta^2 = \Delta_{pg}^2 + \Delta_{sc}^2$, and $f(E_{\textbf{p}})$ is the Fermi function. We define
$c_{\eta} =1/15\pi^2$, $c_{\sigma} =1/3\pi^2$ and the
``coherence factors" are
$u^2_{\bf p},v^2_{\bf p}=\frac{1}{2}(1\pm\frac{\xi}{E})$.
The transport lifetime
$\tau_{\eta,\sigma}$ scales inversely with $\gamma$.
This reduces to the usual BCS results when $\Delta_{pg}^2 = 0$.
Eq.~(\ref{eq:19}) contains two strong correlation effects which
reduce the size of $\eta$ and $\sigma$: the 
presence of a gap $\Delta(T)$ in $E$, which persists into
the normal state and the presence of the
factor $-\Delta_{pg}^2$ which reflects the reduction in the number
of fermions contributing to dc transport as a result of their
conversion to bosons. While the gap is fermionic
in nature, this term is associated with bosonic degrees of freedom.
In the stronger
dissipation limit, the shear viscosity and conductivity can be rewritten 
in terms of generalized Green's functions (to be defined more
precisely in Sec. \ref{sec:3}). 
\begin{widetext}
\begin{eqnarray}
\eta&\!\!=\!\!&-\lim_{\omega\rightarrow0}\lim_{q\rightarrow0}\frac{1}
{\omega}\textrm{Im}\sum_P
\frac{2p^2_xp^2_y}{m^2}\Big[G_{P^+}G_{P^-}-F_{\textrm{sc},P^+}F_{\textrm{sc},P^-}
-F_{\textrm{pg},P^+}F_{\textrm{pg},P^-}
-2(u^2_{\mathbf{p}}-v^2_{\mathbf{p}})^2F_{\textrm{pg},P^+}
F_{\textrm{pg},P^-}\Big]_{i\Omega_l\rightarrow\omega^+}, 
\nonumber \\
\sigma&=&-\lim_{\omega\rightarrow0}\lim_{q\rightarrow0}\frac{1}{\omega}
\textrm{Im}\sum_
P\frac{2p^2_z}{m^2}\Big[G_{P^+}G_{P^-}+F_{\textrm{sc},P^+}F_{\textrm{sc},P^-}
+F_{\textrm{pg},P^+}
F_{\textrm{pg},P^-}
-2(u^2_{\mathbf{p}}+v^2_{\mathbf{p}})^2F_{\textrm{pg},P^+}F_{\textrm{pg},P^-}\Big]_{i\Omega_l\rightarrow\omega^+}
\label{eq:40},
\end{eqnarray}
\end{widetext}
where $P^+=(i(\omega_n+\Omega_l),\mathbf{p}+\frac{\mathbf{q}}{2})$, 
$P^-=(i\omega_n, 
\mathbf{p}-\frac{\mathbf{q}}{2})$, and $\omega^+=\omega+i0^+$. 
Note that while one can interpret $F_{sc}$ as the usual Gor'kov Greens
function reflecting superconducting order, there must also be
a counterpart $F_{pg}$ (discussed in detail later) which reflects non-condensed
pairs. 
In previous work on both cold gases \cite{Bruunvisc2} and cuprates 
\cite{Smith,Levchenko} only the first term involving $GG$ was included,
and, moreover,
this was correctly recognized \cite{Levchenko,Bruunvisc2} as inadequate.

\subsubsection{Summary of Transport Results}
\label{sec:1C2}

We next present summary numerical
plots and compare the viscosity and $d$-wave conductivity calculations
based on Eqs.~(\ref{eq:40}). For definiteness we take a simple
Fermi liquid $\gamma(T) \propto T^2$ temperature dependence for both transport
properties. 
An essential change from $s$-wave to $d$-wave pairing is associated
with the fact that for the latter the carrier number is no
longer exponentially activated. As a consequence, the transport
behavior tends to be more metallic.

The four top panels in Fig.~\ref{fig:13} plot (a) the
$T_c$-normalized, $\omega \rightarrow 0$,  $d$-wave conductivity, 
$\textrm{Re}\sigma(T/T_c)$. In
Fig.~\ref{fig:13}
(b) we plot a decomposition of this conductivity with
three different contributions as color coded, so that
the condensate ($sc$) is blue and the $pg$ contributions are red,
and what makes up the difference is shown in green.. 
Plotted in
Fig.\ref{fig:13}(c) is the
effective carrier number
$(n/m(T))_{\rm{eff}}$
which is obtained by multiplying $\sigma$ by $\gamma(T)$
corresponding to the inverse lifetime.
Finally in Figure \ref{fig:13}(d) we plot the resistivity,
or inverse conductivity, above $T_c$.
The lower panel of figures corresponds to the counterpart
plots for the $s$-wave viscosity,
in terms of $\alpha$ defined via 
$\eta \equiv \alpha
n\hbar$.

The figures in the (b) column show that the pg contributions
are negative in both cases while the condensate gives a positive
contribution to $\sigma$ and
a negative contribution for $\eta$; this can be
directly seen from
Eqs.~(\ref{eq:40}).
The remainder (associated with
the addition of the $GG$ terms in
Eqs.~(\ref{eq:40})),
which yields the respective totals, 
is shown in green.
That the ($pg$) contribution from the
non-condensed pairs (red) lowers the conductivity and viscosity is because
the presence of
non-condensed pairs means fewer fermions are available for
\textit{dc} transport.
Plotted in
in the (c) column one sees that
the effective carrier number
$(n/m(T))_{\rm{eff}}$
increases more or less monotonically as temperature is raised.
This represents a generic figure, since here the temperature
dependence of the lifetime has essentially been removed.
Thus, this figure shows that the effective carrier number associated
with the conductivity and
its shear viscosity counterpart are
increasing functions of temperature,
\textit{strongly suppressed by both the presence of a fermionic
gap and the presence of bosonic degrees of freedom}.
In Fig.\ref{fig:13}(d) we plot the inverse transport coefficients.

The first two figures (top and bottom) on the left in
Fig.\ref{fig:13}
seem to capture the qualitative experimental transport
features shown in Figures
\ref{fig:3} for the cuprates
and
Figure \ref{fig:2} for the cold gases.
The conductivity
exhibits \cite{Orenstein}
a maximum below $T_c$, as can be seen experimentally from Fig.\ref{fig:3},
while the viscosity coefficient monotonically increases,
as observed
\cite{ThomasViscosityScience_online}.
As shown in panel (d), the conductivity 
of the normal state is appropriately metallic,
but
suppressed by the excitation gap.
Importantly, the resistivity has a nearly linear temperature
dependence as observed experimentally, in Fig.~\ref{fig:3}.

For the cold gases, more
direct comparison with experiment involves inclusion of
the trap. As is conventional \cite{Ourreview}, we
include
trap effects via the local density approximation (LDA),
or, equivalently Thomas-Fermi approximation.
Figure~\ref{fig:12}
presents a comparison of the viscosity coefficient
$\alpha$ between theory (based on the RF-deduced lifetime), as black dots, and 
experiment 
\cite{ThomasJLTP08} (red triangles)
as a function
of $E$.
These calculations can also be compared
with more recent experiments summarized in 
Fig. \ref{fig:2}.
Our calculations, which predated the latest experimental
data, will be discussed in more detail later.
One can see, however, that the trends are compatible.

\section{Theoretical Framework of BCS-BEC Crossover}
\label{sec:2}
\subsection{T-matrix Derivation of BCS Theory}
\label{sec:2A}

In order to understand how to
address BCS-BEC crossover at finite $T$, we
now rederive standard BCS theory from a T-matrix scheme. Important here
is that
BCS theory can be viewed as
incorporating \textit{virtual} non-condensed pairs.
Here we consider the general case applicable to both
$s$ and $d$-wave pairing by incorporating
$\varphi_{{\bf
k}}=[\cos(k_x)-\cos(k_y)]$ for the latter and taking
it to be unity for the former.
%
%
%\begin{widetext}
These virtual $ Q \neq 0$ pairs are associated with
an effective
propagator or t-matrix which is constrained to be of the form
\begin{equation}
t(Q) \equiv \frac {U} { 1 + U \sum_{K} G_K G_{0,-K+Q}\varphi_{{\bf
k-q/2}}^{2}}. \label{eq:t}
\end{equation}
in order to yield the standard BCS equations. This t-matrix is associated with
a summation of ladder diagrams in the
particle-particle channel [see Fig. \ref{fig:6} (a)] and
importantly depends on
both 
$G$ and $G_0$, which represent dressed and non-interacting Green's
functions, respectively. We use $K,Q$ to denote four-vectors with
$K=(i\omega_n,{\bf k})$, $Q=(i\Omega_l,{\bf q})$, and where
$\omega_n$ and $\Omega_l$ are fermion and boson Matsubara
frequencies, respectively. In order to describe pairing in the
$d_{x^2-y^2}$-wave channel, we write the attractive fermion-fermion interaction
in the form $U_{{\bf k},{\bf k}^{\prime}}=U\varphi_{{\bf
k}}\varphi_{{\bf k}^{\prime}}$, where $U$ is the strength of the
pairing interaction.
%%%%%%
As in bosonic theories, non-condensed pair excitations of the
condensate are necessarily gapless below $T_c$. This means that
$t(Q\rightarrow0)\rightarrow \infty$  which is equivalent to the vanishing of
the effective pair chemical potential, $\mu_{pair} =0$, for $T \leq T_c$. Thus
we have a central constraint on the $T$-matrix
\begin{equation}
t^{-1}(Q\rightarrow0) = 0 \rightarrow~ \mu_{pair} =0, T \leq T_c \label{eq:t0}
\end{equation}
In order to show that the above condition is identical to the BCS gap
equation, we need the appropriate form for $G_K$. In the BCS theory
the fermionic self energy that appears in the fully dressed Green's
function, $G_K$, is of the form
\begin{eqnarray}
\Sigma_{sc,K} &=& \sum_Q t_{sc}(Q) G_{0,-K+Q} \varphi_{{\bf k}-{\bf q}/2}^{2} 
=-\sum_Q \Delta_{sc,{\bf k}}^2 \delta(Q) G_{0,-K+Q}
= -\Delta_{sc,{\bf k}}^2 G_{0,-K}
\label{eq:Sigsc}
\end{eqnarray}
where $\Delta_{sc,{\bf k}}(T)\equiv\Delta_{sc}(T)\varphi_{{\bf k}}$
is the superconducting order parameter. The full Green's function is
then obtained via the Dyson equation,
$G_K=[G_{0,K}^{-1}-\Sigma_{sc,K}]^{-1}$, which, when inserted in
Eq.~(\ref{eq:t}) yields the BCS gap equation below $T_c$
\begin{equation}
1 = - U \sum_{\bf k}
\frac{1 - 2 f (E_{\bf k}^{sc})} { 2 E_{\bf k}^{sc}}\varphi_{\bf k}^{2}
\label{eq:pureBCSgeq}
\end{equation}
%\end{widetext}
with
$E_{\bf k}^{sc} \equiv \sqrt{\xi_{\bf k}^2 + \Delta_{sc,{\bf k}}^2}$.
We have, thus, used Eqs.~(\ref{eq:t}) 
and
Eq.~(\ref{eq:t0})
to derive the standard BCS gap
equation within a $T$-matrix language and the result appears in
Eq.~(\ref{eq:pureBCSgeq}). Equation~(\ref{eq:t0}) above can be viewed as
representing an extended version of the Thouless criterion of strict
BCS which applies for all $ T \leq T_c$. This derivation leads us to
reaffirm the well known result \cite{Kadanoff,Patton1971,Abrahams}
that BCS theory is associated with one bare and one dressed Green's
function in the pair propagator.

\subsection{Generalization to BCS-BEC crossover}\label{sec:equations}
\label{sec:2B}

To address BCS-BEC crossover, we presume that the
non-condensed ($Q \neq 0$) pairs are no longer virtual.
Thus the T-matrix of
Eq.~(\ref{eq:t}) in general possesses two contributions: the ${\bf
q}=0$ contribution that gives rise to the formation of the condensed
or superconducting pairs and the ${\bf q} \not =0$ contribution that
describes the correlations associated with the non-condensed pairs.
As a result, the fermionic self-energy also possesses two
contributions which are given by
\begin{widetext}
\begin{equation}
\Sigma(K) = \sum_{Q} t(Q) G_{0,-K+Q} \varphi_{{\bf k}-{\bf q}/2}^{2} = \sum_Q [t_{sc}(Q) + t_{pg}(Q) ]
G_{0,-K+Q} \varphi_{{\bf k}-{\bf q}/2}^{2} = \Sigma_{sc,K} + \Sigma_{pg,K}
\label{eq:Sigtotal}
\end{equation}
\end{widetext}
The resulting full Green's function,
$G^{-1}=G_0^{-1}-\Sigma_{sc}-\Sigma_{pg}$ is illustrated in
Fig.~\ref{fig:6}(b). While as before, $\Sigma_{sc,K} =
-\Delta_{sc,{\bf k}}^2 G_{0,-K}$, we find numerically
\cite{Maly1,Maly2} that $\Sigma_{pg,K}$ is in general of the form
\begin{equation}
\Sigma_{pg,K} \approx \frac{\Delta_{pg,{\bf
k}}^2}{\omega+\xi_{\bf k}+i\gamma} \label{eq:4a}
\end{equation}
with $\Delta_{pg,{\bf k}} = \Delta_{pg} \varphi_{\bf k}$. That is,
the self-energy associated with the non-condensed pairs possesses
the same structure as its BCS counterparts, albeit with
a finite lifetime, $\gamma^{-1}$. Physically this arises from
the fact that $t_{pg}(Q)$ is strongly peaked around $Q =0$
below $T_c$ where the pair chemical potential is zero and for
a range of temperatures above $T_c$ as well where this chemical
potential is small.

Analytic self consistent equations for $\Delta_{pg}$ and
$\Delta_{sc}$ can be obtained
microscopically when we consider the small
$\gamma$ limit where
\begin{equation}
\Sigma(K) \approx - (\Delta_{sc,{\bf k}}^2 + \Delta_{pg,{\bf k}}^2)
G_{0,-K} \equiv - \Delta_{\bf k}^2 G_{0,-K} \label{eq:approxSig}
\end{equation}
with  
\begin{equation}\label{eq:pgeq}
\Delta_{pg}^2 \equiv -\sum_Q t_{pg}(Q)
\end{equation}
Here, we have used
the fact that because of the vanishing of $\mu_{pair}$ below $T_c$,
the bulk of the contribution to $\Sigma_{pg}$ in the ordered state
comes from small $Q$. This then leads to an effective pairing gap
$\Delta(T)$ whose square is associated with the sum of the squares
of the condensed and non-condensed contributions
$$\Delta_{\bf k}^2(T)=\Delta_{sc,{\bf k}}^2(T) + \Delta_{pg,{\bf k}}^2(T)$$
Note
that the full gap $\Delta_{\textbf{k}}$ remains relatively T-independent, even below $T_c$, as
observed, because of the conversion of non-condensed ($\Delta_{pg,\textbf{k}}$) to condensed
($\Delta_{sc,\textbf{k}}$) pairs as the temperature is lowered.

The gap equation for this pairing gap,
$\Delta_{\textbf{k}}(T)=\Delta(T)\varphi_{\textbf{k}}$, is again
obtained from the condition $t_{pg}^{-1}(Q = 0)=0$, and given by
\begin{equation}\label{eq:BCSgap}
1 = - U \sum \frac{1 - 2 f(E_{\bf k})}{2 E_{\bf k}} \varphi_{{\bf
k}}^{2},
\end{equation}
This analysis can be made more explicit
after analytical continuation so that
\cite{MalyThesis}, $t_{pg}(\omega,{\bf q})\approx [Z(\Omega -
\Omega^0_{\bf q}+\mu_{pair}) + i \Gamma^{}_Q]^{-1}$, where
$Z=(\partial\chi/\partial\Omega)|_{\Omega=0,q=0}$, $\Omega^0_{\bf
q}\approx q^2/(2M_{b})$ with the effective pair mass
$M_{b}^{-1}=(1/2Z)(\partial^{2}\chi/\partial
q^{2})|_{\Omega=0,q=0}$. Near $T_c$, $\Gamma^{}_Q \rightarrow 0$
faster than $q^2$ as $q\rightarrow 0$ and will be neglected.
Then $\Delta_{pg}^2\approx Z^{-1}\sum_{\bf q}b(\Omega_{\bf q}^{0}-\mu_{pair})$.

Note that one needs to
self-consistently determine the chemical potential, $\mu$, by
conserving the number of particles, $n = 2 \sum_K G_K$, which
leads to
\begin{equation}
n = 2 \sum_KG_K = \sum _{\bf k} \left[ 1 -\frac{\xi_{\bf
k}}{E_{\bf k}} +2\frac{\xi_{\bf k}}{E_{\bf k}}f(E_{\bf k})  \right]
\label{eq:neq}
\end{equation}
Eqs.~(\ref{eq:pgeq}), (\ref{eq:BCSgap}), and (\ref{eq:neq}) present a closed set of
equations for the chemical potential $\mu$, the pairing gap
$\Delta_{\textbf{k}}(T)=\Delta(T)\varphi_{\textbf{k}}$, the pseudogap
$\Delta_{pg,{\bf k}}(T)\equiv\Delta_{pg}(T)\varphi_{{\bf k}}$, and 
the superconducting order
parameter $\Delta_{sc,\textbf{k}}(T)=\Delta_{sc}\varphi_{\textbf{k}}$
with $\Delta_{sc}(T)=\sqrt{\Delta^2(T)-\Delta_{pg}^2(T)}$.  Following this
approximation, $\Delta_{pg}(T)$ essentially vanishes in the ground
state where $\Delta = \Delta_{sc}$. This is to be expected from the
BCS-Leggett wavefunction in Eq.~(\ref{eq:GS}). In this way, the "two
gap" physics disappears in the ground state. Importantly,
numerical studies 
\cite{Chienlattice2} show that 
for $d$-wave pairing, there is no superfluid phase in the
bosonic regime where $\mu$ is negative; the pseudogap
is, thus, associated with the
fermionic regime.

\section{Transport Theory and Gauge invariant approaches to superconductivity and superfluidity}
\label{sec:3}

Our transport theory for BCS-BEC crossover is based on linear response theory 
for both the density
(or charge, labelled $C$) and spin (labelled $S$) degrees of freedom. In this approach, the U(1) electromagnetic
(EM) gauge symmetry and the spin rotational symmetry around the $z$ axis 
play important roles and an understanding of transport in strongly correlated superfluids
has to incorporate in a central way the related 
conservation constraints. 
These enter via
(i) the transverse f-sum rule. Application of the latter to the conductivity
is, in turn, related to (ii) the absence (above $T_c$) and presence (below $T_c$) of a Meissner
effect.

The perturbing Hamiltonian 
can be written in a compact form $\int d^3{\mathbf{r}}(\lambda^{C,S}_{\sigma,\sigma^{\prime}}\psi^{\dagger}_{\sigma}\psi_{\sigma^{\prime}}+\textrm{h.c.})$ where $\lambda^{C,S}_{\sigma,\sigma^{\prime}}\propto\delta_{\sigma\sigma^{\prime}} $ and $\lambda^{S}_{\sigma\sigma^{\prime}} \propto
\delta_{\sigma\bar{\sigma}^{\prime}} $. 
Here $\psi^{\dagger}$($\psi$)
are the fermionic creation (annihilation) operators,
$\sigma=\uparrow$ or $\downarrow$, $\uparrow=-\downarrow$ and
$\bar{\sigma}=-\sigma$. We represent the density- density, current-current and
spin correlation functions as, $\chi_{\rho\rho}$, $\tensor{\chi}_{JJ}$ and 
$\chi_{SS}$.
Experimentally, the last of these three
can be probed by spin-preserving and spin flip (two photon) Bragg scattering. 
The shear ($\eta$) and bulk viscosities ($\zeta_2$)
and the conductivity  
may be written in terms of
the longitudinal 
($\chi_L$)
and transverse 
($\chi_T$)
components of the current-current correlation functions
$\tensor{\chi}_{JJ}$

\begin{eqnarray}
\eta&=&-m^2\lim_{\omega\rightarrow0}\lim_{\mathbf{q}\rightarrow0}\frac{\omega}{q^2}\textrm{Im}\chi_T(\omega,\mathbf{q}), \label{eta0} \\
\zeta_2+\frac{4}{3}\eta&=&-m^2\lim_{\omega\rightarrow0}\lim_{\mathbf{q}
\rightarrow0}\frac{\omega}{q^2}\textrm{Im}\chi_L(\mathbf{q},\omega), 
\label{SFzeta0}\\
\sigma (\omega \rightarrow 0) &=&-\lim_{\omega\rightarrow0}\lim_{q\rightarrow0}\frac{\textrm{Im}\chi_T(\omega,\mathbf{q})}{\omega}
\end{eqnarray}
where the longitudinal
$\chi_L=\hat{\mathbf{q}}\cdot\tensor{\chi}_{JJ}\cdot\hat{\mathbf{q}}$ and
transverse
$\chi_T=(\sum_{\alpha=x}^z\chi^{\alpha\alpha}_{JJ}-\chi_L)/2$ susceptibilities
satisfy \cite{KadanoffMartin}
\begin{eqnarray}
\lim_{\mathbf{q}\rightarrow0}\int_{-\infty}^{+\infty}\frac{d\omega}{\pi}\big(-\frac{\textrm{Im}\chi_T(\omega,\mathbf{q})}{\omega}\big)&=&\frac{n_n(T)}{m},
\label{rulechiT}\\
\int^{+\infty}_{-\infty}\frac{d\omega}{\pi}\big(-\frac{\textrm{Im}\chi_L(\omega,\mathbf{q})}{\omega}\big)&=&\frac{n}{m}. \label{rulechiL} \\
\int_{-\infty}^{\infty}d\Omega Re \sigma(\Omega)&=&e^2\frac{n}{m}
\label{rulesigmaa}
\end{eqnarray}
Because we simultaneously discuss both neutral and charged systems,
it is useful to define $e \equiv 1$ for the neutral case.
Here $n_n$ is the particle number of the normal component in the superfluid
and the sum rules are explicitly
written for the Fermi gas in which there
are no bandstructure effects.
%This second sum rule holds for all temperatures and all wave-vectors, whereas the first
%sum rule is restricted to small wave-vectors; moreover, the right hand side is temperature
%dependent, except above $T_c$ when $n_n \rightarrow n$.
%
For scattering probes, we define the associated structure factors for spin 
and charge or density in
terms of closely related response functions,
$\chi_{\rho\rho}$ and $\chi_{SS}$:  
$$S_{C}(\mathbf{q},\omega)=-\frac{1}{\pi}\textrm{coth}
(\frac{\omega}{2T})\textrm{Im}\chi_{\rho\rho}(\mathbf{q},\omega) \quad
\textrm{and} \quad
S_{S}(\mathbf{q},\omega)=-\frac{1}{\pi}\textrm{coth}(\frac{\omega}{2T})
\textrm{Im}\chi_{SS}(\mathbf{q},\omega)$$
Since the conservation laws for particle number and spin, 
$\partial^{\mu} J_{\mu}=0$ and $\partial^{\mu} J^S_{\mu}=0$ are satisfied,  
the following two sum rules must be 
respected at all temperatures in the whole BCS-BEC crossover regime:
\begin{equation}
\int^{\infty}_{-\infty}d\omega\omega S_{C,S}(\mathbf{q},\omega)= 
\frac{n\mathbf{q}^2}{m}
\quad \textrm{with} \quad \lim_{\mathbf{q}\rightarrow\mathbf{0}}S_{C,S}(\mathbf{q},\omega)=0. 
\label{eq:5}
\end{equation}

The EM kernel is defined by ${\bf {J}} = - \tensor{K}\cdot \bf{A}$,
where 
 $\tensor{K}(Q) = e^2\big(\tensor{n}/m\big)_{dia} + \tensor{P}(Q)$, and the paramagnetic
contribution, given by $\tensor{P}(Q)$, is associated with the normal current resulting from
fermionic and bosonic excitations \cite{Kosztin2,OurBraggPRL}. 
In the superfluid phase,
the density correlations which enter into scattering and the current correlations
which enter into transport can be schematically written
as a sum of 3 terms where
for convenience we drop the ($\omega,\bf{q}$)
arguments:
$\tensor{\chi}_{JJ}=\tensor{P}+\frac{\tensor{n}}{m}+ \overleftrightarrow{\textrm{Coll}}_J$ and
$\chi_{\rho \rho} = P^{00} + \textrm{Coll}_{\rho}$
The counterparts for the spin degrees of freedom are
$\tensor{\chi}_{SS}  =\tensor{Q}_S + \frac{\tensor{n}}{m}$
and
$\chi_{SS} = Q^{00}_S$.
Here $\tensor{P}, ~P^{00}$ and $\tensor{Q}_S,~Q^{00}_S$ represent the ``bare''
contributions.
Collective mode effects in the charge response, which are not
present in the spin response, must also be included
\textit{in the longitudinal response} below $T_c$.
These appear in the above equations as
$\overleftrightarrow{\textrm{Coll}}_J$
and
$\textrm{Coll}_{\rho}$.
These collective mode effects are essential for insuring that the
sum rules and related conservation laws are satisfied.

\begin{figure*}
\includegraphics[width=2.0in,clip]
{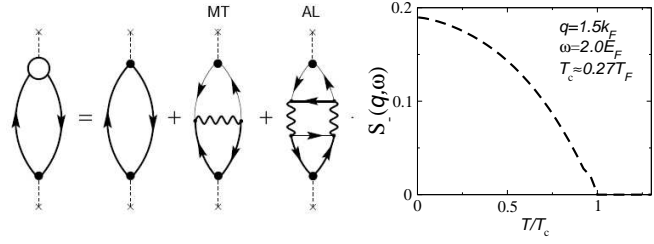}
\includegraphics[width=1.5in,clip]
{S-_q_omega_a0_q1.5.eps}
\caption{Characteristic diagrams which enter into the spin and charge
response kernels (left). Heavy and light lines correspond to dressed and bare
Green's functions. 
Here the wavy lines represent the pair propagator $t$.
The corrections to the vertex include the Maki Thompson (MT)
and Aslamazov-Larkin (AL) diagrams. 
Temperature dependence of $S_-(\mathbf{q})$ (see Eq.~(\ref{eq:8})
in units of $n/2E_F$) for fixed 
momentum
transfer $q=1.5k_F$ and frequency $\omega=2.0E_F$ at unitarity (right) . This figure 
shows that $S_-$ behaves
like an order parameter \cite{OurBraggPRL}.
}
\label{fig:6}
\end{figure*}

In the
most general case, the 
diamagnetic current is expressed in terms of
the inverse band mass $\partial^2\xi_{\textbf{k}}/\partial k_{\alpha}\partial k_{\beta}$ (with $\alpha,\beta=x,y,z$),
via $(\tensor{n}/m)_{dia}=2\sum_{K,\alpha}(\partial^2\xi_{\textbf{k}}/\partial \textbf{k}\partial\textbf{k})G_K$.
\textit{Importantly, the latter
contribution which is temperature independent, should not
be confused with $(n/m(T))_{\rm{eff}}$}. We stress that this
effective carrier number is sensitive to the pairing
gap $\Delta(T)$, while the diamagnetic contribution is not.
We integrate the expression for the diamagnetic contribution by parts and use the
self energy equation and the Generalized Ward identity
to obtain (See Appendix A) an alternate form above $T_c$
\begin{eqnarray}
\Big(\frac{\tensor{n}}{m}\Big)_{dia}\!\!=\!\!-2\displaystyle{\sum_K}\frac{\partial\xi_{\textbf{k}}}{\partial\textbf{k}}\frac{\partial\xi_{\textbf{k}}}{\partial\textbf{k}}\Big[G_K^2
\!+\!\displaystyle{\sum_P}t_{pg}(P)G^2_{0,P-K}G^2_K\Big]\
%n &=& -\frac{2}{3}
%\sum_K \frac{k^2}{m} \big[ G^2(K) %- \Delta_{sc}^2 G_o^2(-K) G^2(K) %\nonumber \\ %&+& \sum_P
%t_{pg}(P) G_o^2(P-K) G^2(K) \big].
\label{eq:1}
\end{eqnarray}
This
exact t-matrix based equation is significant because
it has cast the diamagnetic response in the form of a two particle response function.
That there is no Meissner effect in the normal state is related to a precise
cancellation between the diamagnetic and paramagnetic terms. Noting
$\tensor{P}(0)=-e^2(\tensor{n}/m)_{dia}$, we may
extend $\tensor{P}(0)$ to finite $Q$ to infer
\begin{widetext}
\begin{equation}
{\tensor{P}}(Q)= {2e^2}\sum_K\frac{\partial\xi_{\textbf{k}+\textbf{q}/2}}
{\partial\textbf{k}}\frac{\partial\xi_{\textbf{k}+\textbf{q}/2}}
{\partial\textbf{k}}\Big[G_KG_{K+Q}
+\sum_Pt_{pg}(P)G_{0,P-K-Q}G_{0,P-K}
G_{K+Q}G_K\Big].
\label{eq:20a}
\end{equation} 
\end{widetext}
%
%Here we have added a contribution from the condensate through the introduction of
%generalized Gor'kov functions, $F_{sc}$. These must be determined
%in a consistent fashion which, as in BCS theory, involves one dressed
%and one bare Green's function.
Thus far, our discussion has been quite general, and we have
circumvented any discussion of specific transport diagrams
by building in the absence of a Meissner effect above $T_c$.
One can alternatively \cite{Kosztin2,Chen2,OurBraggPRL}
introduce the Aslamazov-Larkin (AL) and Maki-Thompson (MT) 
diagrams to arrive at the above equation (see Fig \ref{fig:6}), 
but the former which involves two
factors of $t_{pg}$, at first sight, appears more complicated.

Collective mode effects are 
not present in the viscosity and conductivity, because they
represent transverse probes. However, they play an
important role in the density-density response. In this regard, it 
is convenient to define
\begin{equation}
S_{\pm}=(S_C\pm S_S)/2
\quad \textrm{from~ which~ it~ follows~ that} \quad 
\int^{\infty}_0d\omega\omega S_-(\omega,\mathbf{q})=0. 
\label{eq:8}
\end{equation}
This latter is a very unusual sum rule. However, it \textit{must}
be satisfied in any
consistent theory of superfluidity, providing spin and charge are conserved.
In Appendix B we show how this sum rule is satisfied above
$T_c$ for an alternate BCS-BEC crossover theory introduced by
Nozieres and Schmitt-Rink (NSR) \cite{NSR}.

There are other physical 
consequences which can be deduced once one has a conservation-law
consistent theory. The conductivity and the 
shear viscosity 
can alternatively
be written in terms of the bare response $P(Q)$ so that
\begin{equation}
\sigma(\omega)=-\lim_{\mathbf{q}\rightarrow0}\frac{\textrm{Im}P^{xx}(\omega,\mathbf{q})}
{\omega}
\quad \textrm{with} \quad 
\eta=-m^2\lim_{\omega\rightarrow0}\lim_{q\rightarrow0}\frac{\omega}{q^2}\textrm{Im}
P^{xx}(\omega,\mathbf{q}).
\label{eq:9}
\end{equation}
Moreover, in this way, one can see how closely related they are.
We will not in this paper address the bulk viscosity,
principally because we do not have the same level of theoretical control
to satisfy the longitudinal $f$-sum rule, which is also
more problematic below $T_c$, where the Goldstone
bosons appear \cite{Nambu60}. 

\subsection{Weak Dissipation Regime}
\label{sec:3A}

In the previous section we have set up the general linearized response theory.
We now discuss the detailed implementation of this formalism based on 
Eq.~(\ref{eq:selfenergy}). Consistency with conservation laws requires
that transport diagrams assume a specific form, reflecting
the behavior of the self energy. We begin with the so-called
``weak dissipation" regime, where in
Eq.~(\ref{eq:selfenergy})
we presume the quantity $\gamma \approx 0$.
The strong dissipation regime is discussed in
the next subsection.
We have found that for the viscosity there is very little
difference between weak and strong dissipation-based calculations,
but this is not true for the $d$-wave conductivity,
because of the easy excitations of fermions around the nodes.
We indicate in
Figure \ref{fig:6} the characteristic diagrams which enter into the 
generalized 
correlation functions.
As a result
one can show \cite{Ourviscosity}
that the ``bare'' contributions to generalized charge and spin susceptibilities take on an extremely simple form
\begin{widetext}
\begin{eqnarray}
\begin{array}{l}
P^{00}(\omega,\mathbf{q})\\
Q^{00}_S(\omega,\mathbf{q})
     \end{array}
   \Bigg\}
=
\sum_{\mathbf{p}}\Big[\frac{E^+_{\mathbf{p}}
+E^-_{\mathbf{p}}}{E^+_{\mathbf{p}}E^-_{\mathbf{p}}}\frac{E^+_{\mathbf{p}}
E^-_{\mathbf{p}}-\xi^+_{\mathbf{p}}\xi^-_{\mathbf{p}} \pm \Delta^2_{\textrm{sc}}
- \Delta^2_{\textrm{pg}}}{\omega^2-(E^+_{\mathbf{p}}+E^-_{\mathbf{p}})^2}
\big(1-f(E^+_{\mathbf{p}})-f(E^-_{\mathbf{p}})\big)\nonumber\\
\quad-\frac{E^+_{\mathbf{p}}-E^-_{\mathbf{p}}}{E^+_{\mathbf{p}}E^-_{\mathbf{p}}}
\frac{E^+_{\mathbf{p}}E^-_{\mathbf{p}}+\xi^+_{\mathbf{p}}\xi^-_{\mathbf{p}}
\mp \Delta^2_{\textrm{sc}} + \Delta^2_{\textrm{pg}}}{\omega^2-(E^+_{\mathbf{p}}-
E^-_{\mathbf{p}})^2}\big(f(E^+_{\mathbf{p}})-f(E^-_{\mathbf{p}})\big)\Big],
\label{eq:17}
\end{eqnarray}
\end{widetext}
%\begin{eqnarray}
%\overleftrightarrow{P}(\omega,\mathbf{q})&=&\sum_{\mathbf{p}}\frac{\mathbf{p}}{m}
%\frac{\mathbf{p}}{m}\Big[\frac{E^+_{\mathbf{p}}+E^-_{\mathbf{p}}}{E^+_{\mathbf{p}}
%E^-_{\mathbf{p}}}\frac{E^+_{\mathbf{p}}E^-_{\mathbf{p}}-\xi^+_{\mathbf{p}}
%\xi^-_{\mathbf{p}}-(\Delta^2_{\textrm{sc}}-\Delta^2_{\textrm{pg}})}
%{\omega^2-(E^+_{\mathbf{p}}+E^-_{\mathbf{p}})^2}\big(1-f(E^+_{\mathbf{p}})-
%f(E^-_{\mathbf{p}})\big)\nonumber\\
%& &\qquad\quad-\frac{E^+_{\mathbf{p}}-E^-_{\mathbf{p}}}{E^+_{\mathbf{p}}
%E^-_{\mathbf{p}}}\frac{E^+_{\mathbf{p}}E^-_{\mathbf{p}}+\xi^+_{\mathbf{p}}
%\xi^-_{\mathbf{p}}+\Delta^2_{\textrm{sc}}-\Delta^2_{\textrm{pg}}}{\omega^2-
%(E^+_{\mathbf{p}}-E^-_{\mathbf{p}})^2}\big(f(E^+_{\mathbf{p}})-f(E^-_{\mathbf{p}})\big)\Big].
%\end{eqnarray}

\begin{widetext}
\begin{eqnarray}
\begin{array}{l}
\tensor{P} (\omega,\mathbf{q})\\
\tensor{Q}_S (\omega,\mathbf{q})
     \end{array}
   \Bigg\}
=
\sum_{\mathbf{p}}\frac{\mathbf{p}}{m}  
\frac{\mathbf{p}}{m}\Big[\frac{E^+_{\mathbf{p}}
+E^-_{\mathbf{p}}}{E^+_{\mathbf{p}} 
E^-_{\mathbf{p}}}\frac{E^+_{\mathbf{p}}E^-_{\mathbf{p}}-\xi^+_{\mathbf{p}}      
\xi^-_{\mathbf{p}} \mp\Delta^2_{\textrm{sc}}+\Delta^2_{\textrm{pg}}}             
{\omega^2-(E^+_{\mathbf{p}}+E^-_{\mathbf{p}})^2}\big(1-f(E^+_{\mathbf{p}})-     
f(E^-_{\mathbf{p}})\big)\nonumber\\                                             
\qquad\quad-\frac{E^+_{\mathbf{p}}-E^-_{\mathbf{p}}}{E^+_{\mathbf{p}}        
E^-_{\mathbf{p}}}\frac{E^+_{\mathbf{p}}E^-_{\mathbf{p}}+\xi^+_{\mathbf{p}}      
\xi^-_{\mathbf{p}} \pm\Delta^2_{\textrm{sc}}-\Delta^2_{\textrm{pg}}}{\omega^2-    
(E^+_{\mathbf{p}}-E^-_{\mathbf{p}})^2}\big(f(E^+_{\mathbf{p}})-f(E^-_{\mathbf{p}})\big)\Big].
\label{eq:18}
\end{eqnarray}                            
\end{widetext}
where $\omega$ implicitly has a small imaginary part and we define the quantities $E^{\pm}_{\textbf{p}}=E_{\textbf{p}\pm\textbf{q}/2}$ and $\xi^{\pm}_{\textbf{p}}=\xi_{\textbf{p}\pm\textbf{q}/2}$. 
Note that
Eqs.~(\ref{eq:17}) and (\ref{eq:18}) 
are exactly the same as their BCS
counterparts when $\Delta_{pg}^2 \rightarrow 0$.
\textit{It is quite remarkable that when we compare these two equations
we see that 
the charge response functions reflect the \textit{difference} 
$\Delta_{sc}^2 - \Delta_{pg}^2$ while the spin response functions depend on
the total pairing gap
$\Delta^2 = \Delta_{sc}^2 + \Delta_{pg}^2$.
We may say that spin correlation functions know only about pairing whereas their charge
counterparts reflect also coherent superfluid order}.

>From the definition of $\chi_T(\omega,\mathbf{q})$
the shear viscosity is given by
\begin{eqnarray}
\eta&=&-m^2\lim_{\omega\rightarrow0}\lim_{\mathbf{q}\rightarrow0}\frac{\pi\omega}{2q^2}\sum_{\mathbf{p}}\frac{p^2\textrm{sin}^2\theta}{m^2}\Big[\big(1-f(E^+_{\bf p})-f(E^-_{\bf p})\big)\frac{E^+_{\bf p}E^-_{\bf p}-\xi^+_{\bf p}\xi^-_{\bf p}-\delta\Delta^2}{2E^+_{\bf p}E^-_{\bf p}}\big(\delta_1(\omega)-\delta_1(-\omega)\big)\nonumber\\
& &\qquad\qquad\qquad\qquad\qquad\qquad\quad-\big(f(E^+_{\bf p})-f(E^-_{\bf p})\big)
\frac{E^+_{\bf p}E^-_{\bf p}+\xi^+_{\bf p}\xi^-_{\bf p}+\delta\Delta^2}{2E^+_{\bf p}E^-_{\bf p}}\big(\delta_2(\omega)-\delta_2(-\omega)\big)\Big],
\label{eq:7}
\end{eqnarray}
where we have used the
abbreviated notation: $\delta_1(\omega)=\delta(\omega-E^+_{\bf p}-E^-_{\bf p})$, $\delta_2(\omega)=\delta(\omega-E^+_{\bf p}+E^-_{\bf p})$.
Physically, two types of terms appear in the above
equation. Both are well known in standard BCS theory.
The first refers to processes which require a minimal frequency of the
order of $2 \Delta(T)$; they arise from the contribution of fermions
which are effectively liberated by
the breaking of pairs. The second of these
terms, involving $\delta_2$, arises from scattering of fermionic
quasi-particles and is the only surviving contribution to the viscosities,
which are defined in the
$\omega\rightarrow0$ limit.
Note that both contributions involve the \textit{difference} of the
condensed and non-condensed components
($\delta\Delta^2=\Delta^2_{\textrm{sc}}-\Delta^2_{\textrm{pg}}$) with
opposite overall signs.
The low $\omega$ quasi-particle scattering processes are
reduced by the presence of non-condensed pairs -- because they are
associated with a reduction
in the number of fermions. By contrast in the
high $\omega \approx 2 \Delta $ limit
the number of contributing fermions will be increased
by breaking pairs.
We next take the low $\omega, q$ limits in
Eq.~(\ref{eq:7}) and
introduce lifetime (or dissipation) effects by writing delta functions
as Lorentzians
$\delta(\omega\pm\mathbf{q}\cdot\nabla E_{\mathbf{p}}) \rightarrow
\frac{1}{\pi}\frac{\gamma_{\eta}}{(\omega\pm\mathbf{q}\cdot\nabla E_{\mathbf{p}})^2+
\gamma^2_{\eta}}$.
In this way we
arrive at Eq.~(\ref{eq:19}) which was presented in the summary section. Note that
$\eta$ assumes a form similar
to a stress tensor- stress tensor correlation
function.

The transport expressions summarized earlier in Eq.~(\ref{eq:19})
correspond to the generally familiar BCS form 
\cite{Viscositypapers,Dorfle80}
except for the effects associated with
non-zero $\Delta_{\textrm{pg}}$ which appears as
a prefactor $1 - \frac{\Delta_{\textrm{pg}}^2}{E^2}$.
This deviation from unity can be traced to the AL diagrams and hence
may be interpreted as extra ``bosonic'' contributions which
contribute to the normal fluid $n_n$.
Their presence
is physically required for the sum rule
consistency of $\chi_T$.
Note that these terms enter with an overall
reduction coherence-effect-prefactor associated
with
$\xi^2/E^2$.
The negative sign for this bosonic term comes from the fact that
only fermions contribute directly to transport within
a BCS-Leggett-based scheme; the more pairs which are
present, the lower the fermionic contribution to the viscosity.
We see
from Eq.~(\ref{eq:40}) that
$\sigma$ and $\eta$ are alike except
for momentum power law prefactors and coherence factors,
When extended to finite frequency we note that 
unlike in BCS theory, there is a pair creation and pair-breaking
contribution to the optical conductivity. This term is
necessarily absent in a clean BCS superconductor, but may well be
the origin of the widely observed \cite{TimuskRMP}
 ``mid-infrared" contribution to
the finite $\omega$ conductivity.

\subsection{Explicit Proof of the transverse f Sum Rule}
\label{sec:3B}

An important check on our microscopic scheme is to show that it satisfies the sum rule
for the transverse component.
The sum rule which are going to prove is in Eq.~(\ref{rulechiT}).
Our proof here is explicitly for the weak dissipation form of the response
functions and for the Fermi gas, where there are
no bandstructure effects. We generalize this later in the paper.
 
The total particle number is

\begin{eqnarray}
n&=&\sum_{\mathbf{p}}\Big(1-\frac{\xi_{\mathbf{p}}}{E_{\mathbf{p}}}\big(1-2f(E_{\mathbf{p}})\big)\Big)  
=\frac{2}{m}\sum_{\mathbf{p}}\frac{p^2}{3E^2_{\mathbf{p}}}\Big(\frac{1-2f(E_{\mathbf{p}})}{2E_{\mathbf{p}}}\Delta^2-\xi^2_{\mathbf{p}}\frac{\partial f(E_{\mathbf{p}})}{\partial E_{\mathbf{p}}}\Big).
\end{eqnarray}

The superfluid density at general temperature \cite{Kosztin1} is given by
\begin{eqnarray}
n_s=\frac{2}{3}\frac{\Delta^2_{\textrm{sc}}}{m}\sum_{\mathbf{p}}\frac{p^2}{E^2_{\mathbf{p}}}\Big(\frac{1-2f(E_{\mathbf{p}})}{2E_{\mathbf{p}}}+\frac{\partial f(E_{\mathbf{p}})}{\partial E_{\mathbf{p}}}\Big).
\end{eqnarray}

Therefore
\begin{eqnarray}
n_n=n-n_s&=&\frac{2}{3}\sum_{\mathbf{p}}\frac{p^2}{E^2_{\mathbf{p}}}\Big(\frac{\Delta^2_{\textrm{pg}}}{m}\frac{1-2f(E_{\mathbf{p}})}{2E_{\mathbf{p}}}
-\frac{E^2_{\mathbf{p}}-\Delta^2_{\textrm{pg}}}{m}\frac{\partial f(E_{\mathbf{p}})}{\partial E_{\mathbf{p}}}\Big).
\end{eqnarray}

Using
$\tensor{\chi}_{JJ}=\tensor{P}+\frac{\tensor{n}}{m}$ 
leads to
\begin{widetext}
\begin{eqnarray}
& &\lim_{q\rightarrow0}\int_{-\infty}^{+\infty}\frac{d\omega}{\pi}\big(-\frac{\textrm{Im}\chi_T}{\omega}\big)
%=\sum_{\mathbf{p}}\frac{p^2}{6m^2}\Big[\frac{E^2_{\mathbf{p}}-E^2_{\mathbf{p}}+2\Delta^2_{\textrm{pg}}}{E^2_{\mathbf{p}}}\big(\frac{1}{2E_{\mathbf{p}}}-\frac{1}{-2E_{\mathbf{p}}}\big)\big(1-2f(E_{\mathbf{p}})\big)\nonumber\\
%& &\qquad\quad-\frac{E^2_{\mathbf{p}}+E^2_{\mathbf{p}}-2\Delta^2_{\textrm{pg}}}{E^2_{\mathbf{p}}}\lim_{q\rightarrow0}\frac{f(E^+_{\mathbf{p}})-f(E^-_{\mathbf{p}})}{E^+_{\mathbf{p}}-E^-_{\mathbf{p}}}\Big]\nonumber\\
=\frac{2}{3m}\sum_{\mathbf{p}}\frac{p^2}{E^2_{\mathbf{p}}}\Big(\frac{\Delta^2_{\textrm{pg}}}{m}\frac{1-2f(E_{\mathbf{p}})}{2E_{\mathbf{p}}}-\frac{E^2_{\mathbf{p}}-\Delta^2_{\textrm{pg}}}{m}\frac{\partial f(E_{\mathbf{p}})}{\partial E_{\mathbf{p}}}\Big)
=\frac{n_n}{m}.
\end{eqnarray}
\end{widetext}

By contrast, (except in special cases, such as $q \rightarrow 0$,
above $T_c$), the longitudinal sum rule requires numerical proof.
Following
the analysis in Ref.~\onlinecite{OurBraggPRL} we find agreement with
the sum rule to within 5-10\%.
In addition to sum rule consistency,
the appropriate diagram set for computing
all transport properties must be chosen so that $n_s(T)$ vanishes at and
above $T_c$. This is somewhat more complicated \cite{ComparisonReview} to
ensure than in strict BCS theory because there is a finite excitation gap at the
transition. This gap or pseudogap, in turn, reflects the fact that
there are bosonic excitations in addition to the fermions which
deplete the condensate.
In Section \ref{sec:3E} we present a more general argument for establishing
the conductivity sum rule whether the dissipation is weak or strong.

\subsection{Two Photon Bragg Experiments}
\label{sec:3C}

In a recent paper \cite{OurBraggPRL}
we have used this theory to
address the dynamical structure factor and
thereby show how two photon Bragg scattering
can be used
to establish \textit{in situ} the presence of coherent order in a superfluid,
at any temperature, wavevector and frequency. For the most part experiments
on unitary gases have relied on sweeps to the BEC, to find evidence for
condensation. Our analysis is based
on the definitions and sum rule in Eq.~(\ref{eq:8}). It, thus, depends on imposing
the current conservation laws, which have been extensively studied and verified \cite{OurBraggPRL}.
Using
the characteristic diagrams 
shown in Figure \ref{fig:6}, 
which enter into the density-density correlation functions
\textit{we are now led to an important observation: the quantity
$S_-(\omega,\mathbf{q})$
for all $\bf q, \omega, 1/k_Fa$ can be used as an indication of in-situ
superfluid order, without requiring sweeps to the BEC}.
We show a plot of this behavior in the right-hand panel of Figure \ref{fig:6},
where it can
be seen that the difference structure factor vanishes in the normal state. 
An interpretation of this figure is that despite previous claims \cite{RanderiaReview}
\textit{there is no spin-charge separation in the normal or pseudogap state associated
with BCS-BEC crossover}. Spin-charge separation is, however, to be found in the
superfluid phase.
The counterpart normal state calculations can be shown to be valid in
an alternate BCS-BEC crossover theory, based on the Nozieres
Schmitt-Rink scheme \cite{NSR}; this is presented in
Appendix B.

\subsection{Strong Dissipation Approach} 
\label{sec:3D}

We now use the full expression for the self energy
Eq.~(\ref{eq:selfenergy}) to obtain compatible expressions for transport coefficients
The full
Green's function is given by 
\begin{eqnarray} G_K=\Big(i\omega_n-\xi_{\textbf{k}}
+i\gamma-\frac{\Delta_{pg,\textbf{k}}^2}{i\omega_n+\xi_{\textbf{k}}+i\gamma}-\frac{\Delta_{sc,\textbf{k}}
^2}{i\omega_n+\xi_{\textbf{k}}}\Big)^{-1}. 
\end{eqnarray}
where we have added an extra constant term $i \gamma$ in order to
be consistent with the weak dissipation limit in the case that
$\gamma$ becomes small.
To extend 
$\tensor{P}(Q)$
which appears in Eq.~(\ref{eq:20a}) below $T_c$, within a BCS-like formulation
one needs to include terms
of the form $ F_{sc,K}F_{sc,K+Q}$
which represent the usual Gor'kov functions as a product of one dressed
and one bare Green's function ($GG_0$)
\begin{equation}\label{eq:Fsc}
F_{sc,K} \equiv
 -\frac{\Delta_{sc,\textbf{k}}}{i\omega_n+\xi_{\textbf{k}} } \frac{1}{i\omega_n - \xi_{\textbf{k}}
-\frac{\Delta^2_{\textbf{k}}}{i\omega_n+\xi_{\textbf{k}} }}. 
\end{equation}
Here, as before, $\Delta^2_{\textbf{k}} \equiv  \Delta_{pg,\textbf{k}}^2 + \Delta_{sc,\textbf{k}}^2$. 
Then, in the same spirit as our derivation of Eq.~(\ref{eq:4a}) we exploit the fact
that
$t_{pg}(P)$ is strongly
peaked at small $P$, which leads us to approximate
\begin{eqnarray}
\tensor{P}(Q)&\approx&{2e^2}\sum_K\frac{\partial\xi_{\textbf{k}+\textbf{q}/2}}{\partial\textbf{k}}\frac{\partial\xi_{\textbf{k}+\textbf{q}/2}}{\partial\textbf{k}}\Big[G_KG_{K+Q}
+F_{sc,K}F_{sc,K+Q}
-F_{pg,K}F_{pg,K+Q}\Big]\label{eq:fullP}  \\
\label{eq:Fpg}
&\rm{where}& F_{pg,K} \equiv - \frac{\Delta_{pg,\textbf{k}}}{i\omega_n+\xi_{\textbf{k}}+ i\gamma}G_K
\end{eqnarray}
From Eq.~\eqref{eq:fullP} and
$\Re \c \sigma ^{para} (\Omega ) \c  \equiv 
-\Im\c P_{xx}(\Omega)/{\Omega}$ 
the paramagnetic contribution to the dc conductivity 

\begin{eqnarray}\label{eqn:gen_conductivity}
 Re \sigma^{para}(0)\!\!&\approx\!\!-\!\!\displaystyle{\lim_{\textbf{q}\rightarrow0}}\textrm{Im}\displaystyle{\sum_K}\Big[\frac{2e^2}{i\Omega_m}\Big(\frac{\partial\xi_\textbf{k}}{\partial k_x}\Big)^2\Big(G_KG_{K+Q}
\!\!-\!F_{pg,K}F_{pg,K\!+\!Q}\!+\!F_{sc,K}F_{sc,K\!+\!Q}\Big)\Big]_{i\Omega_m\rightarrow0^+}
\end{eqnarray}

In order to be consistent we rewrite Eq.~(\ref{eq:1}), also adding in the usual
BCS condensate terms 
\begin{eqnarray}
\Big(\frac{n_{xx}}{m}\Big)_{dia}\!\!&\approx&\!\!\!-2\displaystyle{\sum_K}\Big(\frac{\partial\xi_{\textbf{k}}}{\partial k_x}\Big)^2\Big[G_KG_{K+Q} 
%\nonumber
-F_{sc,K}F_{sc,K+Q} 
-F_{pg,K}F_{pg,K+Q}\Big]. \label{eq:fullnm} 
\end{eqnarray}
We can use Eq.~(\ref{eq:fullP})
to arrive at Eqs.~(\ref{eq:40}) which were
presented earlier in the form of a summary.
Note that in previous work
in the literature
\cite{Smith,Levchenko,Bruunvisc2} only the first term involving $GG$ was included,
which was recognized \cite{Levchenko,Bruunvisc2} as inadequate.

\subsection{Proof of Conductivity Sum Rule}
\label{sec:3E}

We now revisit the issue of compatibility
with the important conductivity or transverse sum rule 
\begin{equation}
\int_{-\infty}^{\infty}d\Omega \textrm{Re} \sigma(\Omega)=
e^2\big({n_{xx}}/m \big)_{dia}
\label{rulesigma}
\end{equation}
in a more general 
fashion. Here
we now include bandstructure
effects through the effective mass.
Note that we must have two contributions to the conductivity corresponding
to the paramagnetic and diamagnetic terms 
\begin{equation}
\textrm{Re} \sigma(\Omega)
= -{Im P_{xx}(\Omega)}/\Omega + \pi \delta(\Omega) [Re P_{xx}(\Omega) + e^2\big({n_{xx}}/m \big)_{dia}]
\end{equation}
Integrating the first term over frequency we find
$= - \pi \textrm{Re} P_{xx}(0)$, while the
second (delta function) term yields a term
$+ \pi \textrm{Re} P_{xx}(0)$, which leaves only the diamagnetic
contribution and
yields the desired
sum rule.
Note that this analysis holds both below and above $T_c$ and that
\textit{this
sum rule
is intimately connected to the
absence (above $T_c$) and the presence (below $T_c$) of
a Meissner effect}.
Importantly,
since $\big(\frac{n_{xx}}{m}\big)_{dia}$
can be
viewed as essentially independent of temperature, when
there are approximations
in evaluating the transport diagrams, it is appropriate to evaluate
the chemical potential $\mu$ based on the $T$-independence in
Eq.~(\ref{eq:fullnm}).
It should be re-iterated this this diamagnetic contribution is
to be distinguished from
$n_{eff}(T)$, which enters into the dc transport.

\begin{figure*}
\includegraphics[width=5.3in,clip]
{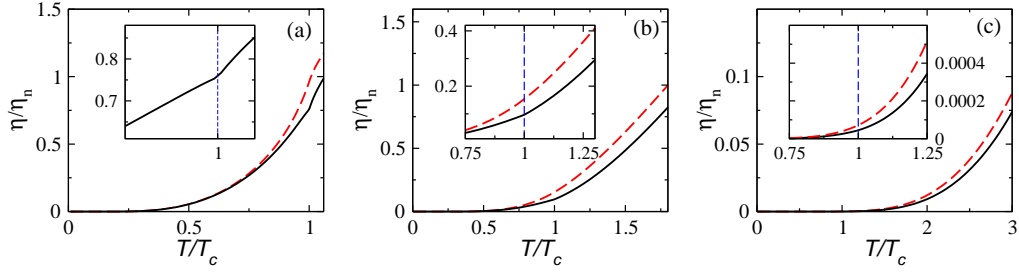}
\caption{(Color online) Theory plots of shear viscosities
as functions of $T$ from BCS to BEC, divided
by the counterpart unpaired state viscosities $\eta_n$.
The red dashed lines are results in absence of
explicit bosonic degrees of freedom
(associated with the AL diagrams). In (a), $T_c=0.12T_F$, $T^{\ast}=0.13T_F$ and $1/k_Fa=-1$. In (b), $T_c=0.26T_F$, $T^{\ast}=0.50T_F$ and $1/k_Fa=0$. In (c), $T_c=0.21T_F$, $T^{\ast}=1.28T_F$ and $1/k_Fa=1$. Expanded plots
in different $T$ regimes are shown in the various insets.}
\label{fig:9}
\end{figure*}

\section{Results for Viscosity}
\label{sec:4}

We now summarize some of our numerical calculations, beginning
with the shear viscosity.
In Figure.~\ref{fig:9} we plot the viscosity divided by 
a ``normal state" value $\eta_n$  
as a function 
of temperature for
$1/k_Fa = \pm 1$ and for unitarity. 
This normal state viscosity is to be associated
with the (temperature dependent) viscosity of the unpaired state
$\eta_n=\int_0^{\infty}dp\frac{p^6}{15\pi^2m^2}
\Big(-\frac{\partial f}{\partial \xi}\Big|_{T^{\ast}}\Big)\tau_{\eta}$.
In this way we take the
same lifetimes in numerator and
denominator for the plots, which necessarily cancel out of the ratio.
For all $k_Fa$, the ratio drops to zero at low temperatures
reflecting the decrease in the number of condensate excitations.
The red dashed line indicates the behavior when the bosonic
excitations are removed and the black solid line shows the
full calculated viscosity. Thus, the shaded regions correspond
to the contribution from the AL diagrams. This contribution is seen to
be largest at unitarity, as expanded in the inset
to 
Figure.~\ref{fig:9}(b).
The ratio $\eta(T_c)/\eta_n$ varies from
0.7 to $4.4\times10^{-5}$
as one passes from BCS to BEC in these three cases. At unitarity the
viscosity at $T_c$ is \textit{reduced by a factor of 10}, relative to
the unpaired fluid.
A  weak signature of the transition is
largest in the BCS regime, as shown in the inset to
Figure.\ref{fig:9}(a).

\begin{figure*}
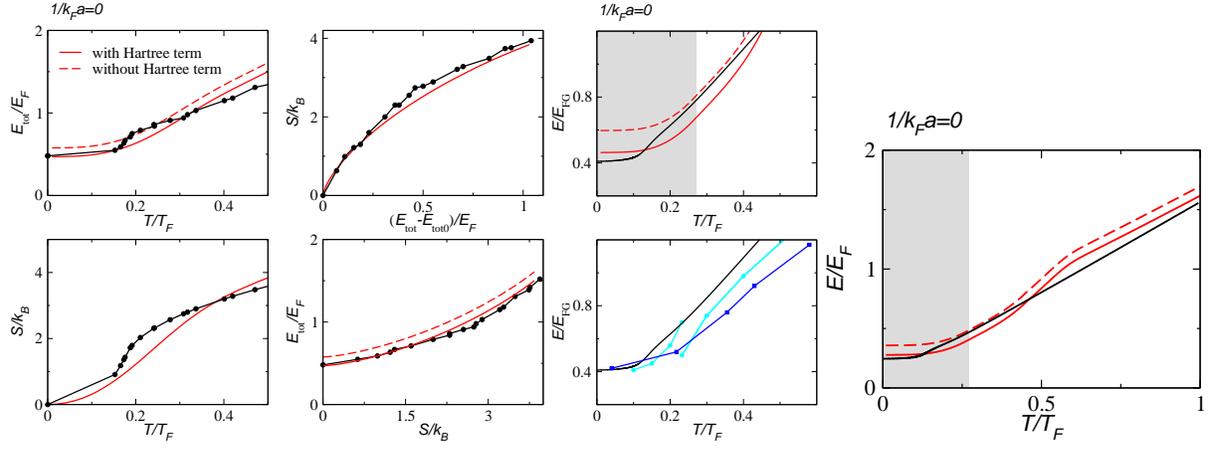

\begin{center}
\includegraphics[width=2.8in,clip]
{OurESa0v2.eps}
\includegraphics[width=1.4in,clip]
{homo_ESa0_H_exp4b.eps}
%{homo_ESa0_H_exp4.eps}
\includegraphics[width=2.in,clip]
{homo_ESa0_H_exp4d.eps}
\caption{Left: Calculated \cite{ThermoScience,ChenThermo,heyan2}
thermodynamical behavior (plotted
as red curves, including Hartree shift (solid) and without (dashed).
Results for the trapped case are shown by 4 left panels and
for homogeneous case by 3 right panels.
Experiments (black lines) on left correspond to those in Ref. \onlinecite{LeLuo}
and those on right
with Ref. \onlinecite{Mukaiyama}.
Third (lower) panel from left 
compares with other
theories, one of which shows the usual spurious first order transition \cite{Drummond2}
and the other represents Monte Carlo simulations \cite{BulgacMC}.
Here $E_{FG}$
is defined as in Ref.~\onlinecite{Drummond2}}.
\label{fig:10}
\end{center}
\end{figure*}

To incorporate trap effects,
we begin by
summarizing past work 
on the thermodynamical properties of cold Fermi gases.
Counterpart thermodynamical experiments have played a role in
characterizing the viscosity \cite{LeLuo,Physicstoday,ThomasJLTP08}.
Figure \ref{fig:10} presents a comparison of
our previous thermodynamical calculations \cite{heyan2,ThermoScience} with
experiments for both the trapped (left) and homogeneous (right) case.
All data is shown in black, while our theoretical
results are shown in red. Agreement is reasonably
satisfactory in both cases, particularly with the addition of a very small Hartree
adjustment which replaces the dashed lines with the solid curves.
The first four panels (top and bottom, from the left) correspond
to comparisons
with the trapped case \cite{LeLuo}. The next three are for
the homogeneous case with the black lines corresponding to
experiments from Japan \cite{Mukaiyama}. The third from the left
two panels compare the present theory (in red) with these experiments
\cite{Mukaiyama} in the upper panel and other analytical (light blue)
\cite{Drummond5} as well as Monte Carlo \cite{BulgacMC} (dark blue)
calculations in the lower panel. The large scale figure shows that the present
approach (red curves) is in quite good agreement with the
homogeneous experiments \cite{Mukaiyama} (in black) over a rather
wide temperature range. Absent here is the first order transition
seen in all other analytic theories.

\begin{figure}
\includegraphics[width=3.0in,clip]
{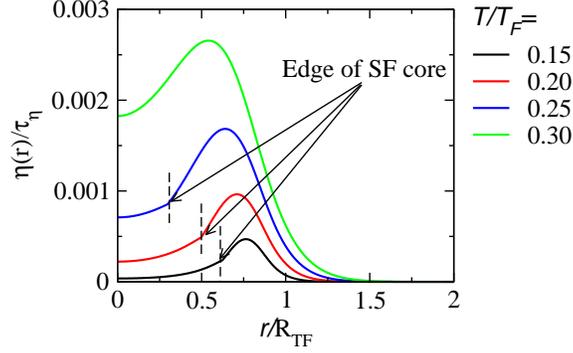}
\caption{
Trap profile for viscosity. The various curves show the shear
viscosity as a function of trap position for four
different indicated temperatures and $R_{TF}$ is the
Thomas-Fermi radius. }
\label{fig:11}
\end{figure}

In Fig. \ref{fig:11} we show the the LDA-calculated viscosity
as a function of position in the trap and for various temperatures $T$.
The arrows indicate the edge of the superfluid (SF) core.
It can be seen that the viscosity is suppressed in the
core region.
In this way
the nearly-free fermions
at the trap edge dominate the trap averaged value for $\eta$.
As a result the calculated viscosity in a trap is considerably
higher than in the homogeneous case.

With these thermodynamical calculations and trap profiles as a backdrop, we now
return to
an earlier figure,
Figure.~\ref{fig:12}, 
which
addressed comparison of theory and experiment for the shear viscosity. For a unitary Fermi gas, we
estimate the transport lifetime from the characteristic
broadening of the
single
particle fermionic spectral function. The latter, in turn, can be found here by
fitting \cite{RFReview,Ourcompare} Radio Frequency (RF) cold gas experiments.
Recall that this fermionic lifetime is associated with
a damping contribution
in an otherwise BCS-like self energy
(conventionally parameterized \cite{MalyThesis,Normanphenom} by ``$\gamma$
'').
The inset to
Figure.~\ref{fig:12}(a) presents a plot of this RF-deduced lifetime as black circles. 
The red triangles in this inset represent a plot of the lifetime which one
would infer from the data for precise agreement between theory and experiment.

Figure.~\ref{fig:12}(a)
presents a comparison of the viscosity coefficient
$\alpha$ between theory (based on the RF-deduced lifetime), as black dots, and experiment \cite{ThomasJLTP08} (red triangles)
as a function
of $E$.
Figure.~\ref{fig:12}
(b) shows the comparison of $\eta/s$ where $s$ is the entropy density.
We find
that $\eta/s$ appears to be relatively $T$ independent at the lower
temperatures.
The last figure on the right represents a transcription 
of the horizontal axis in Figure.~\ref{fig:12}(b) which plots $\eta/s$ 
as a function of temperature, rather than energy $E$.
This is based on using the calculated trap thermodynamics
\cite{ThermoScience} to rescale the various axes. Moreover,
our calculations incorporate 
the same trap averaging procedure as in Ref.~\onlinecite{ThomasJLTP08}.
One can anticipate that, particularly at the lower $T$, the trap-integrated viscosity will be artificially higher
than for the homogeneous case, since $\eta$ will be dominated by
unpaired fermions at the trap edge.
It should be noted that in the homogeneous case
the ratio of the viscosity
to its normal counterpart is exponentially activated with
$T$. This is to be contrasted
with the behavior of the entropy which reflects bosonic
power laws \cite{ThermoScience} in $T$.
Overall, it can be seen that our calculations agree favorably with the 
experimental
data. 
Interestingly, the observed behavior appears more consistent with
previous helium-3 experiments
\cite{Helium3}
than those in helium-4 \cite{Woods}, as can be seen from
Figure \ref{fig:1}.

\section{Conductivity in the cuprates}
\label{sec:5}

It is the goal of this section to address 
the dc conductivity $\sigma(T)$ for a pseudogapped superconductor such
as the high $T_c$ cuprates. We approach this problem in a
fashion which is analogous to our above discussion of viscosity
for an ultracold Fermi gas.
In the cuprate literature one associates a fixed stoichiometry
(hole doping) with a given excitation gap at $T_c$, say. The
lower the hole doping ($x$) the larger this gap. Since the parameter
$x$ is of no particular interest here, we parameterize instead
a given stoichiometry by the size of $\Delta(T_c)$, or alternatively
the size of the temperature $T^*$ at which $\Delta(T)$ first
vanishes. 
The two key puzzles of the dc conductivity 
in the cuprates are the near linearity of the resistivity with
temperature and the fact that only the doped holes
($\sigma (T_c) \propto x)$ appear to contribute to transport.

A central conclusion of our
conductivity study is that, just as for the viscosity,
the reduction in the effective carrier number
$(n/m(T))_{\rm{eff}}$ is revealed to play an important role,
both in the $T$ and $x$ dependence of transport.
This reduction, in turn, is a consequence of the presence
of an excitation gap which persists into
the normal phase, and which increases as $x$ decreases. 
The suppression in the carrier number is substantial
relative to the full
diamagnetic or sum rule value in Eq.~(\ref{rulesigma}).
Moreover,
$(n/m(T))_{\rm{eff}}$ 
rises
nearly monotonically with temperature until $T^*$.
This 
contribution leads to a non-metallic
tendency with $\sigma$ \textit{increasing} with $T$, above $T_c$.
In order to yield a metallic resistivity (which increases with $T$)
the contribution of 
$(n/m(T))_{\rm{eff}}$
must be off-set by taking $\gamma(T)$ to be a higher power than linear.
Here we illustrate our results for $T > T_c$ with
a Fermi liquid like behavior $\gamma(T/T_c)^2$, which could
plausibly be associated with Fermi arcs \cite{Kanigelarcs}, which are extended gapless
regions due to the smearing out of the $d$-wave nodes above $T_c$.
This is the most conventional $T$ dependence for transport
processes which involve inter-fermion scattering. 

We return to Fig.\ref{fig:13} (top panel) which was presented earlier and
which summarizes the general behavior.
Of particular interest is the behavior of
the calculated resistivity which is illustrated in
Fig.\ref{fig:13}d which, itself is roughly linear, because
of the assumed quadratic or Fermi liquid $T^2$ dependence
in $\gamma$.
Note that these conductivity calculations
are specific to the $d$-wave case
and for an $s$-wave counterpart, it would be very difficult to
find metallic behavior. This can be seen by noting
that the inverse viscosity
shown in this figure decreases with increasing $T$, reflecting
the even more strongly suppressed carrier number.

\begin{center}
\begin{figure}
\includegraphics[width=2.2in,clip]
{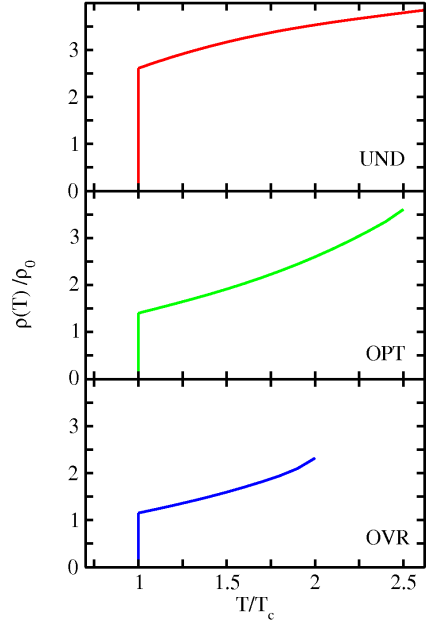}
\caption{Theoretical plots of resistivity as a function of
temperature for three different values of $x$, corresponding
to the pairing onset temperature (or strength of attractive
interaction). The UND, OPT and OVR labels as the
same as those used in Figure \ref{fig:3}, from
Reference \onlinecite{Bontemps}. The resistivity decreases as
doping increases, deriving from an increase in the effective carrier number.
That the plots stop at temperature $T = T^*$ reflect the fact
that Eq.~(\ref{eq:selfenergy}) is no longer appropriate when the system
is this far from condensation.}
\label{fig:14}
\end{figure}
\end{center}

Figure \ref{fig:14}  presents a more detailed
plot of these normal state resistivities 
for dopings that interpolate between the 
heavily underdoped (large $T^*$) and overdoped cases
($T^* \approx T_c$). 
The resistivities are normalized by the value $\rho_o = \rho(T_c)$ for the
case of highest doping. We may
characterize each curve, in order of increasing doping, by the ratio 
$\gamma/\Delta(T_c)=0.07,0.20,~\textrm{and}~ 0.35$. That the
size of the resistivities decreases as one 
increases the hole concentration, largely reflects the change in
gap size and thus in effective carrier number.
Indeed, to a first approximation
$(n/m(T_c))_{\rm{eff}}$
scales close to linearly with $x$.

Although it is subtle, one can see from the figure that
there is a change in the nearly linear slope with increased
doping from concave to convex bowing, which may be
seen experimentally \cite{Bontemps}
in Figure \ref{fig:3}.
In this way, it would appear that the so-called ``optimal" doping
(where the bowing is minimal)
represents a form of mid-way point, rather than a specific
form of ``strange metal" \cite{LeeReview}.

\section{Comparison with Literature}
\label{sec:6}

\subsection{Pseudogap theories}
\label{sec:6A}

Since the present transport calculations emphasize the
role of the pseudogap, it is useful to clarify the nature
of the pseudogap which is specifically associated with BCS-BEC crossover. 
One of the earliest observations that a pseudogap may be
present in BCS-BEC crossover theories is due to Randeria
and co-workers \cite{RanderiaVarenna}. This
first generation analysis 
focused on ``spin gap" effects,
so that there were a number of claims which are inconsistent
with current understanding. Despite statements to
the contrary \cite{RanderiaVarenna}, (i) The pseudogap
is not a ``spin gap", but is also to be associated, 
as we have seen in the present transport calculations, with
the ``charge channel".  This mirrors the experimental
observations \cite{Sutherland} on the nature of
the cuprate pseudogap: ``It is a quasiparticle gap and not
just a spin gap." (ii) The pseudogap is not associated
with ``spin-charge" separation above $T_c$. In this
paper (Section \ref{sec:3E} and Appendix B) we have shown that
the absence of a Meissner effect leads to
the non-separation of spin-charge degrees of freedom.
(iii) The pseudogap in crossover theories is quite
distinct from the pseudogap associated with the phase
fluctuation scenario (as discussed in Section \ref{sec:1B1}).

Our group was one of the earliest (i) To flesh out an understanding
of the nature of this pseudogap, and, in particular
to show \cite{Janko} that the
BCS-BEC crossover-pseudogap is not only associated with spin as was
claimed \cite{RanderiaVarenna}, but that more generally
it represents a quasi-particle
gap. This observation was based on calculations of the
fermionic spectral function. 
(ii)
To address gauge invariant electrodynamics in the presence of a pseudogap
\cite{Kosztin1,Chen2}.
(iii)
To introduce the concept
of a pseudogap into the cold gas literature \cite{JS2} and
(iv) To introduce \cite{MalyThesis} the now widely used
expression \cite{Normanphenom}
for the self energy (Eq.~(\ref{eq:selfenergy})).

\subsection{Cold Fermi Gases}
\label{sec:6B}

There has been a significant effort devoted to reaching
a theoretical understanding
of the various viscosities associated with highly correlated
fermionic superfluids \cite{Son1}. A detailed overview exploring the
relation between cold atomic gases and hot quark gluon plasmas
was presented by Schaefer and Teaney \cite{Schaefer}.
Taylor and Randeria \cite{TaylorRanderia}
established sum rules at unitarity
for the dynamical counterpart $\eta(\omega)$ and $\zeta_2(\omega)$,
which are less general than
those in 
Eqs.~(\ref{rulechiT}) 
and (\ref{rulesigma}).
Zwerger and colleagues \cite{Zwergervisc} addressed the behavior
of the normal unitary gas predicting a minimum in $\eta/s$ near
the superfluid transition.

Bruun and Smith \cite{Bruunvisc1,Bruunvisc2}
were, perhaps, the first to emphasize the importance of
the (static) shear viscosity in cold Fermi gases using both 
a high temperature fermionic
Boltzmann approach and a Kubo-stress tensor based scheme
(above $T_c$),
within BCS-BEC theory.
They importantly recognized \cite{Bruunvisc2}
that the introduction of a pseudogap
would lower the normal state $\eta$. However, the diagram set which was used was
``not conserving''\cite{Bruunvisc2}.

Rupak and Schafer \cite{Rupakvisc}
introduced an alternative (bosonic
Boltzmann) transport theory
in which the low $T<< T_c$ shear viscosity is dominated
by the Goldstone bosons or phonons.
They predicted that $\eta/s$,
increases as temperature decreases.
In their approach phonons dominate the low $T$ transport,
presumably leading to the same upturn in $\eta$, as is
seen in helium-4, and shown in Figure \ref{fig:1}, on the right.
Establishing diagrammatic consistency is a central theme
of this paper and within
a BCS-like theory the shear
viscosity, as a transverse response,
does not directly couple \cite{Nambu60} to the
Nambu-Goldstone
bosons.
This is to be distinguished from the longitudinal current-current
correlations which are the basis for the bulk viscosity and necessarily
depend on these collective modes.

\subsection{Transport in High Tc superconductors}
\label{sec:6C}

Understanding the temperature dependence of the (inverse)
conductivity or dc resistivity, particularly near optimal doping, was
one of the first puzzles posed by the high temperature superconductors. 
It should, nevertheless be noted that a number of ideas and
interpretations about transport
were established well before there was a full understanding
of
how widespread was
the pseudogap phase (which extends over most of the phase
diagram, including
optimal doping). For this reason, one can argue
that these earlier theories must be
reexamined.

A number of 
different
models from spin-charge separation \cite{LeeReview} to marginal 
Fermi liquid phenomenology 
\cite{VarmaLittlewood} were invoked to explain (i) the unusual
normal state
temperature power laws observed, \textit{e.g.,} the
linear resistivity.
A second
notable observation (ii) was that the effective carrier number in transport 
appeared to
scale with the number of (extra) doped holes \cite{LeeReview}, 
called $x$, even though
the
volume of the Fermi surface scaled as expected with $1+x$. 
The first of these experimental observations gave rise to the concept of
a distinct ``strange metal" phase which was thought to separate
the overdoped and underdoped regimes. 
The second of these gave rise to the concept of ``spin-charge"
separation.
Despite the widespread agreement that the normal state of
the cuprates was a non-Fermi liquid phase, there is today
a strong belief that the superconducting phase is
Fermi liquid based \cite{LeeReview}.

It must be noted that the BCS-BEC crossover scenario addresses
these issues from a very different perspective. 
It presumes that there is a
smooth evolution, as seen experimentally \cite{Deutscher2}
from over to underdoped behavior rather than
a distinct strange metal phase near optimal doping. We have
argued here that the suppression in the number of carriers
may be associated with the fact that the magnitude of
the excitation gap grows with underdoping. [In the ac
conductivity, this leads to
a very small weight for the $\omega \approx 0$ Drude peak,
which requires by the sum rule, an additional mid-infrared
contribution, now widely observed experimentally \cite{TimuskRMP}.] 
As noted
above, there is no spin-charge separation in the normal
phase. Importantly,
the superconducting phase has bosonic excitations representing
non-condensed pairs (also
present above $T_c$) and therefore this phase is not
Fermi liquid based.
Finally, the dissipation mechanism here associated with the
BCS-BEC crossover scenario arises from the inter-conversion
of fermions and pairs, which is distinctly tied to
the pseudogap. This is to be distinguished from most
transport theories \cite{Hirschfeld2,Leecond} which focus on impurity effects
or the dynamics of the pairing boson.

There is a shortcoming in our cuprate transport
calculations because we have ignored
impurity effects in transport altogether. Away from very low $T$,
it is generally accepted \cite{TimuskRMP} that they are not particularly
important. Nevertheless, they may lead to incomplete
condensation \cite{Orenstein,Leecond} in the ground state.
Indeed, it is difficult to see how to reconcile claims \cite{TimuskRMP} 
that the $T=0$ superfluid density scales as $ \propto x$, with the
transverse f-sum rule 
(Eq.~\ref{rulesigma})),
without
invoking incomplete condensation in the ground state.
A deeper understanding of a possibly more inhomogeneous
treatment of impurities is needed.
Fortunately, this is not an issue in the cold gases.

With a growing appreciation for the nature of the pseudogap
(and related ``Fermi arc" effects), experimentalists have
provided some support to our findings.
From Ref.\onlinecite{AndoRes1}
it is said that ``This indicates that the functional form of
the dc resistivity of cuprates $\rho_{dc}(T)$  is governed not only
by the relaxation processes 
but also by temperature-dependent numbers of carriers".
Moreover, in
Ref.\onlinecite{AndoRes2} it is stated that
``One may notice that a natural extension of the 
present argument
would be that the T-linear resistivity 
usually observed near optimum doping may
not
necessarily by a sign of a T-linear relaxation rate,
because $n_{\rm{eff}}$ may be changing with T.".
An interesting corollary to a carrier number which
necessary increases with $T$, is that the inverse
lifetime $\gamma$ should contain higher powers than
linear (we use the most conventional, Fermi liquid
dependence $\gamma \propto T^2$) to arrive
at metallic behavior for the conductivity.

\section{Conclusions}
\label{sec:7}

In summary, this paper has addressed the role of the
pseudogap in the $\omega \rightarrow 0$ conductivity
and in the shear viscosity both above and below $T_c$.
We have emphasized the analogy 
between ``bad metals" \cite{EmeryKivelsonPRL74}
and
``perfect fluids" \cite{Physicstoday} seen in high $T_c$
superconductors and the atomic Fermi
gases. Both of these phenomena, we argue, may
arise from pseudogap effects.

Our approach builds on a consistent gauge invariant
treatment of transport (which has not been
addressed previously), in which the transverse f-
sum rule demonstrably
holds. 
In this paper we have demonstrated success in
simultaneously
addressing experiments
in cold gases and high $T_c$ cuprates within the same transport
formalism.
As a summary, Figure \ref{fig:13} gives a reasonable
understanding of the experiments shown in 
Figures \ref{fig:2} and \ref{fig:3}.  
Equally important, Figure \ref{fig:12} shows semi-quantitative
agreement with shear viscosity data from below to above
$T_c$. We see no sign of the upturn which others have
predicted.

It was our intention in this overview to introduce some
of the key challenges in understanding transport in the
high $T_c$ superconductors to the wider readership
interested in perfect fluidity.
It is hoped that an appreciation of this broader context
may lead to new breakthroughs in understanding superficially
distinct,
but quite possibly connected, 
physical systems.

\vskip 10mm

This work is supported by NSF-MRSEC Grant
0820054. We thank Le Luo and John Thomas and T. Mukaiyama, 
for sharing their data and Benjamin M. Fregoso for helpful
conversations.
C.C.C. acknowledges the support of the U.S.
Department of Energy through the LANL/LDRD Program.

\appendix
\section{Rewriting the number equation}
\label{app:A}
It is useful to fill in a few steps in the line
of argumentation from the text. We wish to show
here for the simpler case of a Fermi gas, how the
number equation may be rewritten using a Ward
identity: 
\begin{widetext}
\begin{equation}
P(Q) = \frac{-2 e^2}{3 m^2} \sum_K (k +\frac{q}{2})^2 \big[ G_KG_{K+Q}
~~+ \sum_P t_{pg} (P) G_{0,P-K-Q}G_{0,P-K} G_{K+Q}G_K \big]
\end{equation}
And the number equation
\begin{eqnarray}
n &=& 2 \sum_K G_K= 2 \sum_K \partial k_{\alpha} /\partial k_{\alpha} G_K
\nonumber \\
&=& 
2 \sum_K k_{\alpha}G^2(K) \partial G^{-1}_K / \partial k_{\alpha}
\nonumber \\
&=& 2 \sum_K k_{\alpha} G^2_K \big[ \partial G_{0,K}^{-1}  / \partial k_{\alpha}
- \partial \Sigma(K) / \partial k_{\alpha} \big]
\nonumber \\
&=& -2 \sum_K k_{\alpha} G^2_K [ \frac {k_{\alpha}}{m} - \sum_Q t(Q) \partial G_{0,Q-K}^{-1}  /\partial k_{\alpha}]
\nonumber \\
&=& -2 \sum_K k_{\alpha} G^2_K [ \frac{k_{\alpha}}{m} + \sum_Q t(Q) G_{0,Q-K}^2 \frac{k_{\alpha}- q_{\alpha}}{m}]
\nonumber \\
n &=& -\frac{2}{3} \sum_K \frac{k^2}{m} \big[ G_K^2 + \sum_Q t_{pg}(Q) G_{0,Q-K}^2 G^2_K \big]
\end{eqnarray}

where the last step specializes the result to $T>T_c$. The last equation is the central result we wanted to prove: 
from which we have
\begin{equation}
K(0) = \frac {ne^2}{m} +P(0) = 0,~~~~\rm{above~T_c}
\end{equation}
\end{widetext}

\section{NSR Theory and the Normal state Structure Factor}
\label{app:B }

In a recent Physical Review Letter \cite{OurBraggPRL}
we have shown how two photon Bragg scattering
can be used
to establish \textit{in situ} the presence of coherent order in a superfluid,
at any temperature, wavevector and frequency. For the most part experiments
on unitary gases have relied on sweeps to the BEC, to find evidence for
condensation. Our analysis is based
on the definitions and sum rule in Eq.~(\ref{eq:8}). It, thus, depends on imposing
the current conservation laws, which have been extensively studied and verified \cite{OurBraggPRL}.

As an alternate example, here we consider Nozieres Schmitt-Rink (NSR)
theory \cite{NSR} in the normal state.
Figure \ref{fig:6} indicates the characteristic
class of diagrams.
These diagrams enter into the density-density 
and spin-spin correlation functions.
The calculations for the spin response build
on an earlier publication \cite{Ourcompare}.
We define the pair susceptibility $\chi_o(Q)=\sum_PG_{0\uparrow,P}G_{0\downarrow,Q-P}
$ where $G_{0\sigma}$ is the bare Green's function.
The pair propagator is given by $t_o(Q)=U/(1+U\chi_o(Q))$, where $U$ is the two-body interaction. Here we present a more systematic discussion of
the spin response than what was shown in Sec.\ref{sec:3}.
The interaction term in the Hamiltonian 
is given by $H_I\sim\int d^3{\bf r}J^S_{\mu}A^{S\mu}$ where $A^S_{\mu}=(B_z,\mathbf{m})$ is the ``effective'' 4-vector field and $J^S_{\mu}=(n^S, \mathbf{J}^S)$. Here $B_z$ is the $z$ component of the magnetic field, $\mathbf{m}$ is the magnetizing field, $n^S$ is the $z$ component of spin and $\mathbf{J}^S$ is the magnetization current. The spin rotational symmetry around the $z$ axis leads to the conservation law of spin: $\partial^{\mu}J^S_{\mu}=0$.

From linear response theory the spin
response kernel can be written as
\begin{equation}
Q^S_{\mu\nu}(Q)=\sum_P\lambda^S_{\mu\sigma}(P,P+Q)G_{0\sigma,P+Q}\Lambda^S_{\nu\sigma}
(P+Q,P)G_{0\sigma,P}+\frac{n}{m}g_{\mu\nu}(1-g_{\mu0}),
\label{eq:10}
\end{equation}
where $\lambda^S_{\mu\sigma}(P,P+Q)=\big(\frac{S_{\sigma}}{m}(\mathbf{p}+\frac{\mathbf{q}}{2}),S_{\sigma}\big)$ is the bare vertex function of the spin-external field interaction, and $\Lambda^S_{\mu\sigma}$ is the full vertex function.
There is an implicit summation over the indices $\sigma$. Note, importantly,
that the vertex function has different signs for different spin indices, which is to be contrasted with
the charge or equivalently density response functions.
In order to satisfy local conservation laws, the vertex must satisfy a Ward identity:
$Q\cdot\sum_P\lambda^S_{\mu\sigma}(P,P+Q)=S_{\sigma}\big(G^{-1}_{0\sigma,P}-G^{-1}_{0\sigma,P+Q}\big)$.
As a result, in Nozieres Schmitt-Rink theory there are three contributions
$\Lambda^S_{\sigma}=\lambda^S_{\sigma}+\delta\Lambda^S_{\sigma\textrm{MT}}+\delta\Lambda^S_{\sigma\textrm{AL}}$,
where the subscript MT is associated with the contribution from the Maki-Thompson (MT) like diagrams and AL the Aslamazov-Larkin (AL) diagrams. 
An important result (for singlet pairing)
is that the contribution from the AL diagrams automatically vanishes.
Thus, for
the spin structure factor
$S_S=S_{S0}+S_{\textrm{MT}^S_{\textrm{pg}}}$
with
$S_{\textrm{MT}^S_{\textrm{pg}}}
=
-S_{\textrm{MT}^C_{\textrm{pg}}}$.
\vskip 1mm
By contrast, in the normal state,
the dynamical structure factor in NSR theory,
for the particle density \cite{Ourcompare}
can be
written as the sum
$S_C=S_{C0}+S_{\textrm{MT}^C_{\textrm{pg}}}+2 S_{\textrm{AL}}$.
Here one should note that
$S_{S0}=S_{C0}$.
The remaining terms denote the
corrections from the MT, and AL
diagrams.
Note that the spin and charge contributions from the MT term enter with
opposite signs.
Moreover, the Ward identity implies
a cancellation \cite{Ourreview}
$S_{\textrm{MT}^C_{\textrm{pg}}}+2 S_{\textrm{AL}} =
-S_{\textrm{MT}^C_{\textrm{pg}}}$.
Thus
\begin{equation}
S_S=S_{S0}+S_{\textrm{MT}^S_{\textrm{pg}}}
= S_C=S_{C0}-S_{\textrm{MT}^C_{\textrm{pg}}},
\label{eq:15}
\end{equation}
which proves the desired result: \textit{ in the normal state} the
spin and charge degrees of freedom are indistinguishable.
Note that the theoretical proof of this result depends on using a consistent
theory of BCS-BEC crossover with full gauge invariance. The same diagrams that
are used in the above proof are needed to show that there is no Meissner
effect in the normal state.
In this way we have demonstrated
\begin{equation}
S_-(\omega,\mathbf{q})\equiv 0 ~~~\rm{above~the~transition.}
\label{eq:16}
\end{equation}
\textit{We are now led to an important observation: the quantity
$S_-(\omega,\mathbf{q})$
for all $\bf q, \omega, 1/k_Fa$ can be used as an indication of in-situ
superfluid order, without requiring sweeps to the BEC}.
We show a plot of this behavior in the right side of Figure 
\ref{fig:6}  where it can
be seen that the difference structure factor vanishes in the normal state. The
behavior below $T_c$ is shown in the figure for
the case of BCS-Leggett theory \cite{Ourcompare}.

%\bibliography{Review2.bib}

\end{document}